\documentstyle[pstricks,epsf,float]{mn_let}

\newif\ifAMStwofonts

\def\beginrefs{\begin{thebibliography}{99}
   }
   \def\apj#1#2#3#4{\bibitem{} #4 19#3, { ApJ,\/} { #1}, #2 }
   
   \def\apjs#1#2#3#4{\bibitem{} #4 19#3, { ApJS,\/} { #1}, #2 }
   \def\aj#1#2#3#4{\bibitem{} #4 19#3, AJ, { #1}, #2}
   \def\pasj#1#2#3#4{\bibitem{} #4 19#3, PASJ, #1, #2 }

   \def\aa#1#2#3#4{\bibitem{} #4 19#3, { A\&A,\/} { #1}, #2 }
   
   \def\aas#1#2#3#4{\bibitem{} #4 19#3, { A\&AS,\/} { #1}, #2 }
   
   \def\mnras#1#2#3#4{\bibitem{} #4 19#3, MNRAS, { #1}, #2 }

   \def\apjl#1#2#3#4{\bibitem{} #4 19#3, { ApJ,\/} { #1}, #2 }

   \def\nature#1#2#3#4{\bibitem{} #4 19#3, { Nat,\/} { #1}, #2 }
   
   \def\asr#1#2#3#4{\bibitem{} #4 19#3, { Adv. Space Res.,\/} { #1}, #2 }

   \def\astph#1#2#3{\bibitem{} #3 19#2, { Preprint Astro-ph} No. #1}
   
   \def\BIB {\bibitem{}}
\def\endrefs{\end{thebibliography}}

\def\hea4{{\it HEAO~A4}}
\def\heaoa2{{\it HEAO~A2}}
\def\heao1{{\it HEAO~1}}

\def\amin{$^\prime$}

\def\eg{{\it e.g.}~}

\def\h0{$H_{\rm o}=50$~km~s$^{-1}$~Mpc$^{-1}$}
\def\q0{$q_{\rm o}$}

\def\msun     {$M_{\odot}$}
\def\lsun     {$L_{\odot}$}

\def\etal    {{ et~al.}~}

\def\cms3  {~{cm$^{-3}$}}

\ifoldfss
  \ifCUPmtlplainloaded \else
    \NewTextAlphabet{textbfit} {cmbxti10} {}
    \NewTextAlphabet{textbfss} {cmssbx10} {}
    \NewMathAlphabet{mathbfit} {cmbxti10} {} 
    \NewMathAlphabet{mathbfss} {cmssbx10} {} 
  \fi
  \ifAMStwofonts
    \ifCUPmtlplainloaded \else
      \NewSymbolFont{upmath} {eurm10}
      \NewSymbolFont{AMSa} {msam10}
      \NewMathSymbol{\upi}     {0}{upmath}{19}
      \NewMathSymbol{\umu}     {0}{upmath}{16}
      \NewMathSymbol{\upartial}{0}{upmath}{40}
      \NewMathSymbol{\leqslant}{3}{AMSa}{36}
      \NewMathSymbol{\geqslant}{3}{AMSa}{3E}

    \fi
  \fi
\fi 

\ifnfssone
  \newmathalphabet{\mathit}
  \addtoversion{normal}{\mathit}{cmr}{m}{it}
  \addtoversion{bold}{\mathit}{cmr}{bx}{it}
  \newmathalphabet{\mathbfit} 
  \addtoversion{normal}{\mathbfit}{cmr}{bx}{it}
  \addtoversion{bold}{\mathbfit}{cmr}{bx}{it}
  \newmathalphabet{\mathbfss} 
  \addtoversion{normal}{\mathbfss}{cmss}{bx}{n}
  \addtoversion{bold}{\mathbfss}{cmss}{bx}{n}
  \ifAMStwofonts
    \ifCUPmtlplainloaded \else
      \UseAMStwoboldmath
      \makeatletter
      \new@mathgroup\upmath@group
      \define@mathgroup\mv@normal\upmath@group{eur}{m}{n}
      \define@mathgroup\mv@bold\upmath@group{eur}{b}{n}
      \edef\UPM{\hexnumber\upmath@group}
      \new@mathgroup\amsa@group
      \define@mathgroup\mv@normal\amsa@group{msa}{m}{n}
      \define@mathgroup\mv@bold\amsa@group{msa}{m}{n}
      \edef\AMSa{\hexnumber\amsa@group}
      \makeatother
      \mathchardef\upi="0\UPM19
      \mathchardef\umu="0\UPM16
      \mathchardef\upartial="0\UPM40
      \mathchardef\leqslant="3\AMSa36
      \mathchardef\geqslant="3\AMSa3E
    \fi
  \fi
\fi 

\ifnfsstwo
  \DeclareMathAlphabet{\mathbfit}{OT1}{cmr}{bx}{it}
  \SetMathAlphabet\mathbfit{bold}{OT1}{cmr}{bx}{it}
  \DeclareMathAlphabet{\mathbfss}{OT1}{cmss}{bx}{n}
  \SetMathAlphabet\mathbfss{bold}{OT1}{cmss}{bx}{n}
  \ifAMStwofonts
    \ifCUPmtlplainloaded \else
      \DeclareSymbolFont{UPM}{U}{eur}{m}{n}
      \SetSymbolFont{UPM}{bold}{U}{eur}{b}{n}
      \DeclareSymbolFont{AMSa}{U}{msa}{m}{n}
      \DeclareMathSymbol{\upi}{0}{UPM}{"19}
      \DeclareMathSymbol{\umu}{0}{UPM}{"16}
      \DeclareMathSymbol{\upartial}{0}{UPM}{"40}
      \DeclareMathSymbol{\leqslant}{3}{AMSa}{"36}
      \DeclareMathSymbol{\geqslant}{3}{AMSa}{"3E}
    \fi
  \fi
\fi 

\ifCUPmtlplainloaded \else
  \ifAMStwofonts \else 
    \def\upi{\pi}
    \def\umu{\mu}
    \def\upartial{\partial}
  \fi
\fi

\title[Constraining the Role of SN~Ia and SN~II in Galaxy Groups] 
{Constraining the Role of
SN~Ia and SN~II in Galaxy Groups 
 by Spatially Resolved Analysis of ROSAT and ASCA Observations}

\author[A. Finoguenov and T. J. Ponman]
       {A.~Finoguenov$^{1}$\thanks{E-mail: alexis@hea.iki.rssi.ru}\ and T.J.~Ponman$^{2}$\\
 $^1$Space Research Institute, Profsoyuznaya 84/32, 117810 Moscow, Russia.\\
 $^2$School of Physics \& Astronomy, University of Birmingham, Edgbaston,
Birmingham B15 2TT, UK}
\date{Accepted for MNRAS}
\pagerange{\pageref{firstpage}--\pageref{lastpage}}
\pubyear{1999}

\def\LaTeX{L\kern-.36em\raise.3ex\hbox{a}\kern-.15em
    T\kern-.1667em\lower.7ex\hbox{E}\kern-.125emX}

\begin{document}

\label{firstpage}

\maketitle

\begin{abstract}
We present the results of modelling the distribution of gas properties in
the galaxy groups HCG51, HCG62 and NGC5044, and in the poor cluster AWM7,
using both ASCA SIS and ROSAT data. The spectral quality of the ASCA data
allow the radial distribution in the abundances of several elements to be
resolved. In all systems apart from HCG51, we see both central cooling
flows, and a general decline in metal abundances with radius. The ratio of
iron to alpha-element abundances varies significantly, and in comparison
with theoretical supernova yields, indicates a significant contribution to
the metal abundance of the intergalactic medium (IGM) from type Ia
supernovae. This is seen both within the groups, and also throughout much of
the cluster AWM7. The total energy input into the IGM from supernovae can be
calculated from our results, and is typically 20-40 per cent of the thermal
energy of the gas, mostly from SNe~II. Our results support the idea that the
SN~II ejecta have been more widely distributed in the IGM, probably due to
the action of galaxy winds, and the lower iron mass to light ratio in groups
suggests that some of this enriched gas has been lost altogether from the
shallower potential wells of the smaller systems.

\end{abstract}


\section{Introduction}

In this paper we present a study of the X-ray emission of galaxy groups
using the good spatial resolution of the ROSAT PSPC in conjunction with the
superior spectral capability of the ASCA SIS. For relatively cool
systems like groups ($T_{\rm gas}\sim 1$~keV) the X-ray spectrum
contains fairly strong signatures from a number of elements, and this
allows us to investigate the distribution of heavy elements in the IGM.
Making use of the different element yields of SN~Ia and SN~II, we can use
this information to estimate the role of different types of SNe in
enriching the IGM.

According to standard theories of the chemical evolution of early-type
galaxies, their star formation will trigger a galactic wind, causing a
loss of the enriched gas, which can be retained in the potentials of
groups and clusters of galaxies (\eg Renzini \etal 1993). 
In the first substantial study of cluster abundances using ASCA,
Mushotzky \etal (1996) studied the integrated abundances of 
O, Ne, Si, S and Fe in four galaxy clusters, and concluded that the
balance between $\alpha$-elements and iron favours the origin
of most or all of the metals in SN~II. More recently Fukazawa \etal (1998)
have studied a much larger sample of 40 clusters, and have also
discriminated out the central regions, which may be dominated by
strong cooling flows. Their results indicate a significant contribution
from SN~Ia, in addition to SN~II, especially in lower mass systems.
In principle, ASCA can provide more detailed information on the spatial
distribution of elements, rather than just on integrated abundances, and
in the present study we exploit this capability. Our work complements
that of Fukazawa \etal (1998), in that we derive detailed distributions
for a small sample, where they study the integrated properties of
a large collection of systems.

The distribution of metals in low mass clusters is of particular interest,
since it is for such systems that the energetic effect of galaxy wind
injection should have the greatest impact on the IGM. It
appears that such effects have already been seen in the low beta values
for groups and in a steepening of the $L:T$ relation at low temperatures
(Ponman \etal 1996, Cavaliere, Menci \& Tozzi 1997). These winds bring
with them metals, so that a study of metal distributions places
constraints on the history of wind injection. If, for example, winds have
blown much of the SN~II metals out of groups altogether, we would expect
them to have lower iron mass to light ratios, and abundance ratios tilted
more towards SN~Ia than is the case in rich clusters. A study of the role
of SN~Ia and SN~II in groups may also shed light on the mystery of iron
production by ellipticals. The low abundance observed by ASCA in hot elliptical
galaxy halos causes substantial difficulties for theoretical models of
their chemical history (Arimoto \etal 1996).

For our study, we require good photon statistics, and have therefore
picked three of the brightest X-ray groups. Two of these (HCG51 and HCG62)
are compact groups, whilst the third (NGC5044) is a loose group.
In practice (Mulchaey \etal 1996) the properties of
compact groups appear to be similar to those of (X-ray bright) loose
groups, and the distinction between the two may not be fundamental.
For comparison, we also study a richer
system, AWM7, (which should lie above the L:T break), but one which is still
cool enough to allow ASCA to determine abundances for a range of elements.

In our study we use a 3-dimensional approach to modelling of ASCA data for
these objects, and we also compare our results for HCG62 with an independent
3-dimensional analysis of ROSAT PSPC data using the Birmingham cluster
fitting package.  {\h0} is assumed throughout the paper. Unless clearly
stated otherwise, all errors quoted in this paper correspond to 90 per cent
confidence level.

\section{Analysis}

In the analysis of ASCA data, presented below, we used data from the SIS0
and SIS1 detectors, which provide energy resolution of $\sim$75~eV at 1.5
keV. A detailed description of the ASCA observatory as well as the SIS
detectors can be found in Tanaka, Inoue \& Holt (1994) and Burke
\etal (1991). Standard data screening was carried out using FTOOLS version
3.6.  The effect of the broad ASCA PSF was simulated, as described in
Finoguenov \etal (1998), and the analysis allows for the projection onto the
sky of the three-dimensional distribution of emitting gas. In this approach,
we use ROSAT data to derive the source's surface brightness profile, which
is represented by a one or two component ``$\beta$-model'' (Jones \&
Forman 1984). This is used to construct a three-dimensional model which is
projected onto the regions of the SIS detectors chosen for extracting
spectra. The thickness of the shells in our ``onion peeling'' technique is
chosen to avoid any drastic variation of the temperature ($T$) and
metallicity ($Z$) within one bin, which would invalidate our single
component spectral modelling.

GIS data were not used in our analysis due to their lower energy resolution
and uncertainties in calibration below 2~keV. Our technique for analyzing
the ASCA data restricts us to usage of ASCA pointings with less than
$25$\amin\ offset from the source centre, due to the similar restrictions in
calibration data for the ASCA ray-traced PSF and the importance of 
accounting for the stray light at larger offset angles (cf Ezawa \etal
1997). In practice this restriction is of importance only for the AWM7
analysis.

All fits are based on the $\chi^2$ criterion. No energy rebinning is done,
but a special error calculation was introduced following Churazov \etal
(1996), to avoid ill effects from small numbers of counts. ROSAT (Truemper
1983) images of all the sources were used as an input for ASCA data
modelling.  For imaging analysis of the ROSAT data we used the software
described in Snowden \etal (1994) and references therein. For the ROSAT/PSPC
observation of HCG62 we also performed an independent three-dimensional
modelling analysis using the Birmingham cluster fitting package (Eyles \etal
1991) for comparison with our ASCA results.

In our approach to the analysis of ASCA data, where all the derived spectra
are fitted simultaneously, and where correlation between different regions
is rather high, a simple minimization of $\chi^2$ proves to be an
ill-posed task (Press \etal 1992, p.795). The best fitting model
displays large fluctuations in temperature and metallicity profiles
as it attempts to fit noise in the data. The solution to this problem
is to accept a much smoother model which gives an acceptable fit
(rather than the very best fit) to the data. This is achieved through
regularization, which applies a prejudice in favour of some measure
of smoothness (Press \etal 1992, p.801). We have made a default
assumption that a linear function (of log radius) is a good representation
of our temperature and abundance distributions. Departures from 
linearity are quantified by calculating the second derivative of the
distribution, which is squared and added to the $\chi^2$ statistic.

In using any regularization technique, special attention should be paid
to the balance between maximizing the likelihood and optimizing the
smoothness of the solution. This is traditionally done by introducing a
weighting constant into the regularization term. We set this constant so
as to obtain a solution which lies within the 68 per cent confidence region
surrounding the minimum $\chi^2$ (i.e. unsmoothed) solution in the
n-parameter space. The allowed offset in $\chi^2$ was evaluated using the
prescription of Lampton \etal (1976).

\begin{table}
\begin{center}

{
\centering
\caption{\protect\small
Optical characteristics. }
\label{tab:opt}
\begin{tabular}{lrllllr}
Name &  D  & $L_{B, cD}$ & $L_B^{^a}$      & $R_{L_B}$ & $R_{v}$& $R_{c}$\\ 
  & Mpc &  $10^{11}$\lsun\ & $10^{11}$\lsun &  Mpc      &     Mpc  &      kpc\\
HCG62   & 82  &      & 0.95    & 1.09 & 1.09 &  27\\
HCG51   & 155 &      & 2.8     & 1.19 & 1.19 &  59\\
NGC5044 & 54  & 0.68 & 1.6$^b$ & 0.5  & 1.19 & 180\\
AWM7    & 106 &  1.9 & 13.2$^b$& 1.84 & 1.89 & 220\\
\end{tabular}
}
\vspace{1pc}

\end{center}

{$^a$}{\small ~ blue luminosity of early-type member galaxies}

{$^b$}{\small ~  Assuming 30\% contribution of
spirals to a total blue luminosity}

\end{table}


Since regularization introduces a dependence of the measurement in a given
spatial bin on that from the adjacent bins (due to the operation of the
prejudice towards smoothness), presenting formal error bars at every point is
misleading. Instead, more valid representation is an area of possible
parameter variation with radius. We choose a confidence level of 90 per cent
for these representations. Thus, points will represent the best-fit
solution, while a shaded zone will mark the area of solutions allowable within
the chosen confidence level.

The analysis presented in this paper differs from previous work on these
objects (\eg\ Fukazawa \etal 1996), which have generally ignored the
effects of projection and PSF blurring. Since the PSF is energy-dependent,
simplified analyses can give misleading results (Takahashi \etal 1995).
For relatively cool systems, like the groups studied here, there is an
additional effect connected with the possible presence of temperature
gradients. Determination of iron abundance at
temperatures $\sim1$~keV is based on L-shell lines. Individual lines
in the L-shell complex are unresolved by ASCA, and merge into 
a broad peak with a position which is sensitive to
temperature.  If the temperature structure is not resolved and a mean
temperature is used, then iron abundance will be underestimated by a
significant factor, because there will be some mismatch between peak
positions. 

To illustrate this effect, we simulate an SIS spectrum using the
characteristics of the HCG62 data (ARF file, response file, duration of the
observation). The model involves a mix of 2 temperatures: 0.7~keV and
1.2~keV, with both components having equal emission measures (a ``norm''
parameter of $2.\times10^{-4}$ in XSPEC, typical for the groups analyzed
here) and all element abundances equal to 0.5 solar. MEKAL plasma code is
used throughout this illustration. When the simulated data are fitted with a
single-temperature model we find $kT=0.823\pm0.026$, Mg$=0.18\pm0.14$,
Si$=0.11\pm0.08$ and Fe$=0.19\pm0.03$ with emission measure 
normalization being
$7.3\pm1.6\times10^{-4}$. Hence the abundances are underestimated, and the
emission measure overestimated. Such effects may compromise any study in
which temperature structure is not resolved. This is likely to apply even to
the current study, in central cooling flows, where steep temperature
gradients and multiphase gas may be present.

In all our spectral modelling, we use the MEKAL model (Mewe \etal 1985,
Mewe and Kaastra 1995, Liedahl \etal 1995). Abundances of He and C were
set to solar values as in Anders \& Grevesse (1989), where element number
abundances relative to hydrogen are (85.1, 12.3, 3.8, 3.55, 1.62, 0.36,
0.229, 4.68, 0.179)$\times10^{-5}$ for O,Ne,Mg,Si,S,Ar,Ca,Fe,Ni,
respectively. In the case of the three groups, only Mg, Si and Fe
features are clearly seen in the data. We cut the ASCA spectra above
2.2~keV, so S and Ar features are outside the range of measurements, and
arrange elements into four groups, based on their production history: O
and Ne; Mg; Si, S and Ar; and Ca, Fe and Ni. The analysis yields
meaningful values for abundances of Mg, Si and Fe. O and Ne results are not
derived for groups, and we do not group O with other elements, like Si,
in order to avoid possible effects from ASCA calibration uncertainties at
low energies. For the same reason, we do not present any results for O
abundance in the analysis below.

Our choice of the MEKAL plasma code leads to systematically lower
temperatures (at the 10--20 per cent level) in the temperature range below 1
keV, compared to the Raymond-Smith code (Raymond \etal 1977). There have
been made several attempts to quantify the uncertainty in the Fe abundances
determined using L-shell line emission. It was shown in a study by
Matsushita (1998), that if Fe abundance is decoupled from the other heavy
elements, the range in Fe abundances derived from different codes
is 20 per cent.  We therefore add a
systematic error of $\pm10$ per cent at 1$\sigma$ confidence level to the
derived Fe abundances.

\begin{figure}
\vspace*{-2.5cm}
\centerline{\scalebox{0.5}{\epsfbox{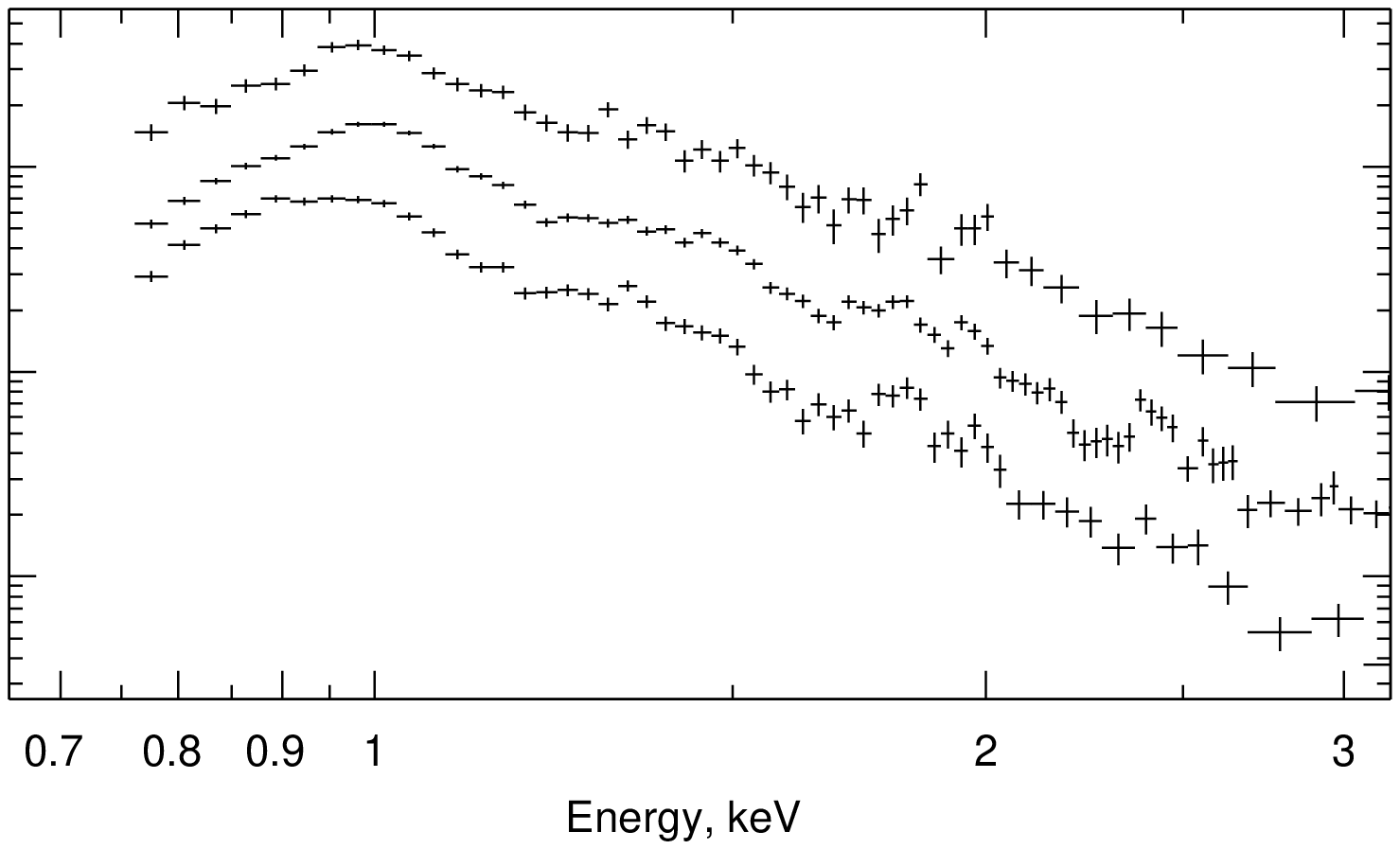}}}
\vspace*{-5.4cm}
\centerline{\scalebox{0.5}{\epsfbox{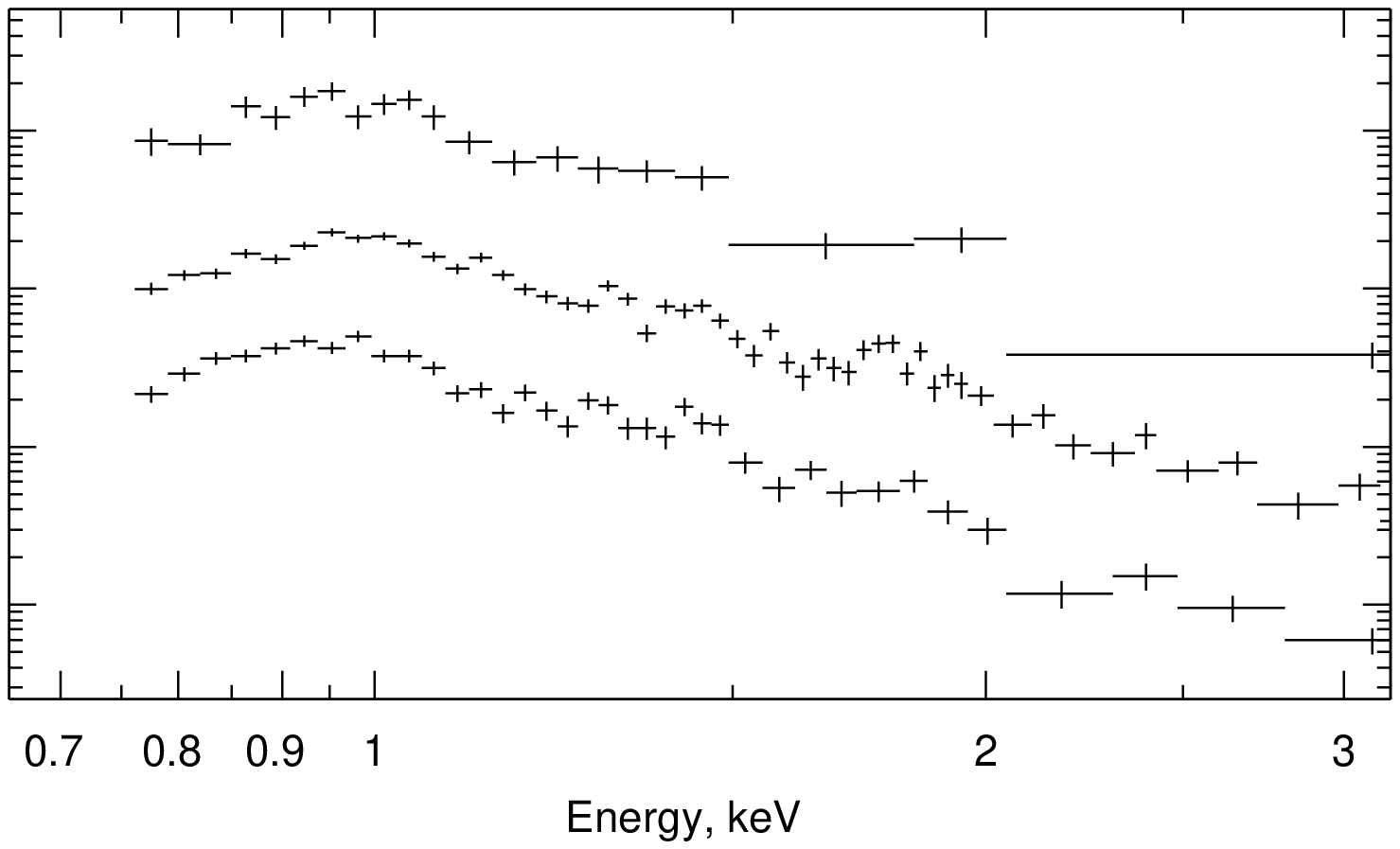}}}
\vspace*{-5.4cm}
\centerline{\scalebox{0.5}{\epsfbox{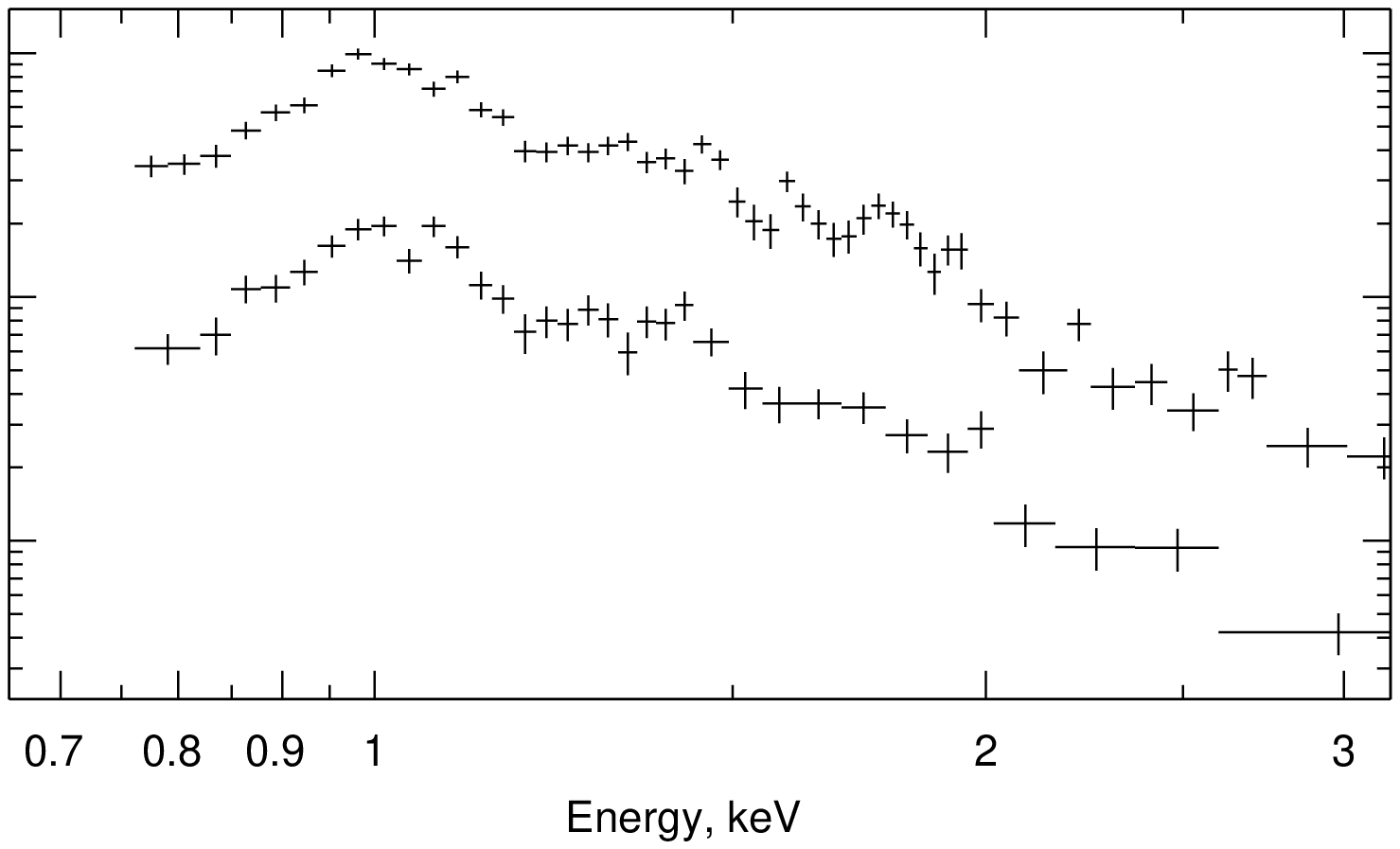}}}
\vspace*{-5.4cm}
\centerline{\scalebox{0.5}{\epsfbox{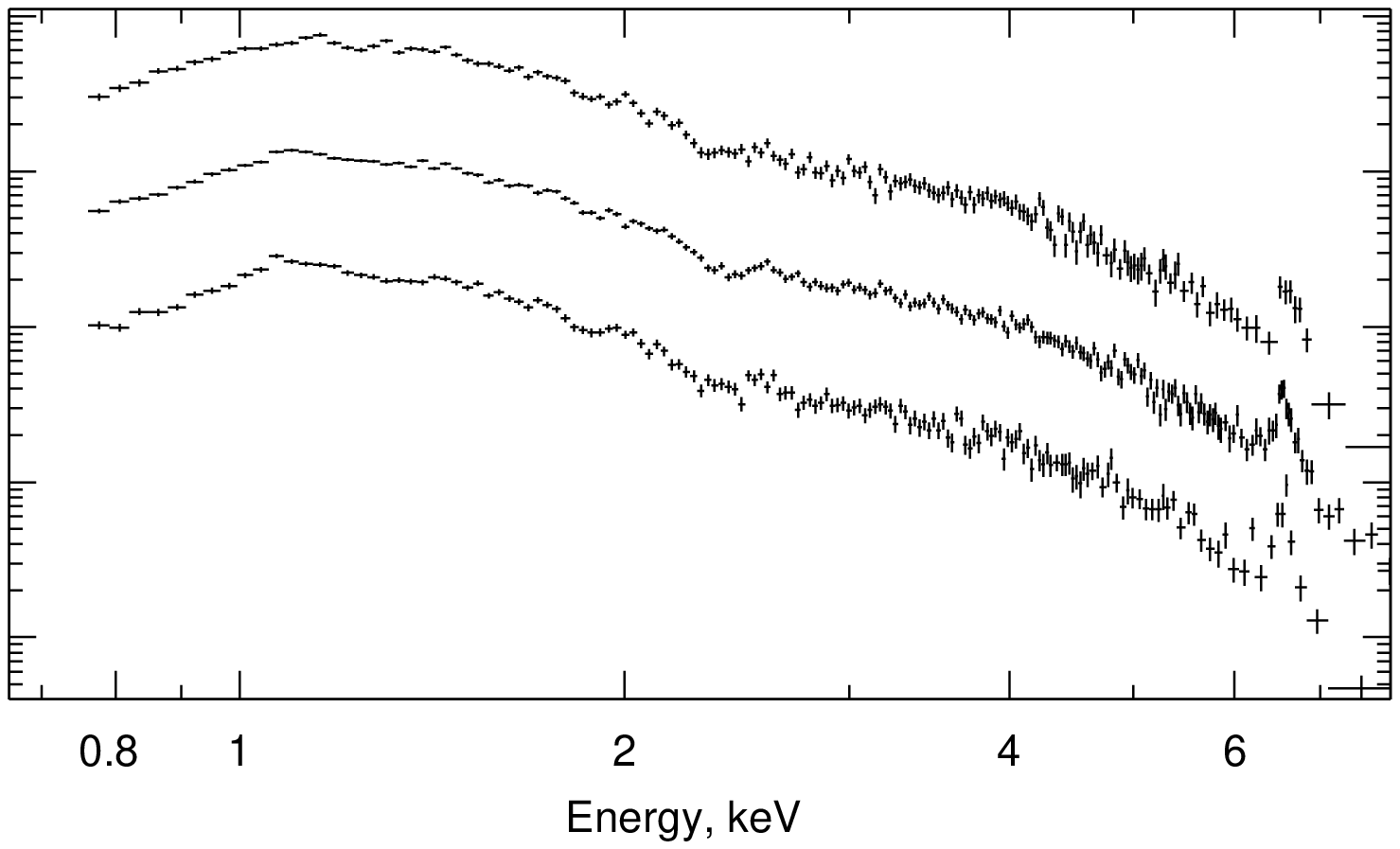}}}
\vspace*{-2.5cm}
\rput(7.4,19.1){{\large\it (a)}}
\rput(7.4,14.4){{\large\it (b)}}
\rput(7.4,9.6){{\large\it  (c)}}
\rput(7.4,4.85){{\large\it (d)}}
\rput(0.5,10.6){\rotateleft{\LARGE\bf S c a l e d \ \ \ f l u x \ \ \  ( c o u 
    n t s \ \ \ s$^{-1}$ \ \ \ k e V$^{-1}$ )}}
\caption{Spatially resolved ASCA SIS spectra of (a) NGC5044, at radii
  of 0\amin--2\amin--8\amin--16\amin\, from bottom to top plots,
  (b) HCG62, at radii 0\amin--1.5\amin--8\amin--16\amin\, from bottom to
  top plots, (c) HCG51, at radii 0\amin--2\amin--6\amin\ from bottom to
  top plots, and (d) AWM7, at radii 0\amin--3\amin--10\amin--25\amin\ 
  from bottom to top plots). Y-axis values are arbitrarily scaled for
  clarity.} 
\label{ex-spe}
\vspace*{-0.5cm}
\end{figure}

Using the 0.7--7.0 keV band in the case of AWM7, we fit all major elements
separately, except for the grouping of Ca and Ni with Fe, and obtain useful
results for the abundances of Ne, Si, S and Fe. A special comment should be
made on Mg.  It has been noted (Mushotzky \etal 1996) that in the case of
cluster emission, this may be affected by the proximity of the poorly
understood 4--2 transition lines of iron, which cannot be resolved with ASCA
SIS.  This problem affects our AWM7 data, however at the lower temperatures
of groups, the relevant Fe L-shell lines are weak and hence uncertainties in
modelling them has little impact on the abundance of Mg determined.
Spatially resolved ASCA SIS spectra obtained for the sources in our study are
illustrated in Fig.\ref{ex-spe}.

\vspace*{-20pt}

\section{Results}

In this section we present the details of the analysis and results for each
individual system analyzed. Optical data, which will be used for comparison
with the X-ray results, are provided in Table \ref{tab:opt}. Columns in this
table are: (1) system name, (2) adopted distance in Mpc, (3) B-band
luminosity of the central galaxy for the two systems with dominant central
galaxies, (4) total luminosity of early-type galaxies, and (5) radius within
which this optical luminosity is measured. Optical data were taken from
Hickson \etal (1992) for HCG51 and HCG62, David \etal (1994) for NGC5044 and
Beers \etal (1984) for AWM7. For the latter object we adopt a spiral
fraction of 0.3, average for clusters, since no measurements are available
to date, however our results are insensitive to this choice. The remaining
columns give: (6) virial radius calculated using a formula derived from
simulations by Navarro \etal (1994), $R_{virial}=1.04 T_{keV}^{0.5}
(H_{o}/50)^{-1}$ Mpc, where T is luminosity weighted X-ray temperature
(excluding the cooling flow), and (7) optical core radius taken from Hickson
\etal (1992), Ferguson \& Sandage (1990) and Dell'Antonio \etal (1995).

\begin{figure}
\vspace*{-0.5cm}
\centerline{\scalebox{0.5}{\epsfbox{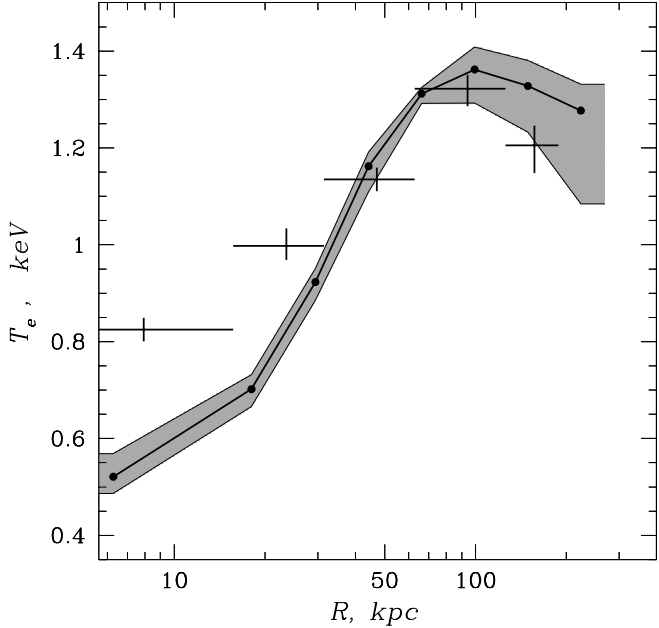}}}
\vspace*{-20pt}
\caption{Temperature profile of NGC5044. ROSAT PSPC
points are represented by black crosses, solid line represents the best-fit
curve describing ASCA results with filled circles indicating the spatial
binning used. Shaded zone around the best fit curve denotes the 90 per cent
confidence area.}
\label{n5044te}
\vspace*{-0.2cm}
\end{figure}

Masses of gas and various metals derived in the X-ray analysis discussed
in the remainder of this section are collected together in 
Table~\ref{tab:mass}. The radius within which these masses are derived is
listed in the second column of the table. All error intervals are quoted at
90 per cent confidence level. Limits on masses are obtained by integrating
the corresponding lower and upper 90 per cent boundary on the abundance
estimation and adding an error on gas mass estimation in squared.

\subsection{NGC5044}

\begin{figure}
\vspace*{-0.5cm}
\centerline{\scalebox{0.5}{\epsfbox{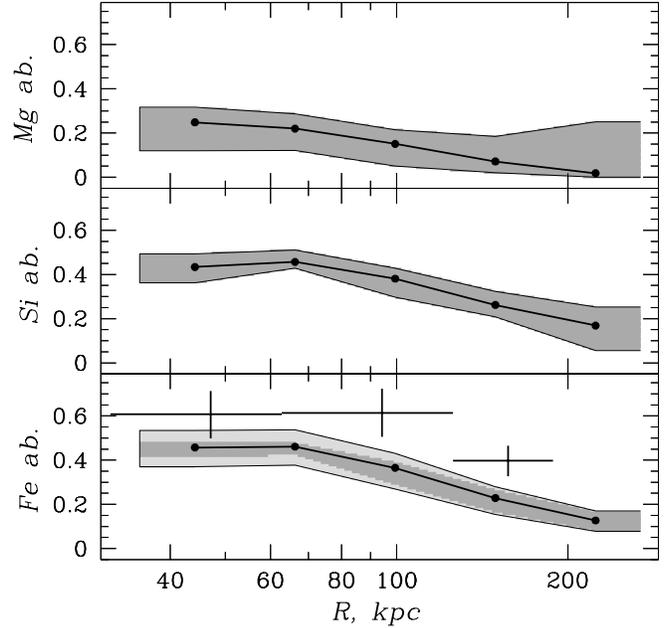}}}
\vspace*{-20pt}
\caption{Element abundances in NGC5044. ROSAT PSPC points are represented by
black crosses, solid line represents the best-fit curve describing ASCA
results with filled circles indicating the spatial binning used. Shaded zone
around the best fit curve denotes the 90 per cent confidence area. Light
shaded zone on the Fe abundance plot indicates the contribution to this zone
of a 10 per cent systematic error, as discussed in the text.}
\label{n5044ab}
\vspace*{-0.2cm}
\end{figure}

\begin{table*}
\begin{center}

{
\centering
\caption{\protect\small
Cumulated gas and metal masses. }

\label{tab:mass}
\begin{tabular}{lrcccccc}
Name &  $R_{out}$ & $M_{gas}$ & $M_{Fe}$ & $M_{Si}$ & $M_{Mg}$ & $M_{S}$ & $M_{Ne}$ \\
     & kpc &  $10^{12}$ \msun & $10^{8}$ \msun & $10^{8}$ \msun & $10^{8}$
     \msun & $10^{8}$ \msun & $10^{8}$ \msun \\
HCG 62   & 410 & 1.656 (1.56--1.76) & 3.82 (1.8--5.8) & 2.24 (0.5--3.4) & 0.23 (0.1--3.8) & --- & --- \\
HCG 51   & 270 & 0.609 (0.56--0.65) & 4.05 (3.0--5.1) & 1.99 (1.6--2.3) & 1.59 (0.6--1.9) & --- & --- \\
NGC 5044 & 270 & 0.965 (0.92--1.00) & 4.09 (2.7--4.9) & 1.75 (1.1--2.2) & 0.49 (0.2--1.5) & --- & --- \\
AWM 7    & 790 & 21.34 (21.1--21.5) & 95.2 (74--115)  & 75.7 (61--89) & --- & 9.4 (1.3--18) & 113 (55--149) \\
\end{tabular}
}

\end{center}

\end{table*}


The galaxy group NGC5044 is well-suited for spatially resolved
spectroscopy, it is nearby ($z=0.0087$), and shows bright and fairly
symmetrical diffuse X-ray emission (David \etal\ 1994). The group has a
dominant central elliptical surrounded by many smaller galaxies. From
analysis of ROSAT PSPC observations (David \etal\ 1994) the diffuse X-ray
emission of this object extends to $\sim20$\amin\ radius (1\amin\
corresponds to 15.7~kpc). A strong temperature gradient is observed in
the centre of the diffuse emission, pointing to the presence of a cooling
flow.  Deprojection gives a central temperature $\sim$0.8~keV. The mass
accretion rate is significant only within the central 40~kpc and amounts
to $\sim10$\msun\ per year. The gas cooling time at $r\approx40$~kpc is
$10^9$~yr.

ASCA observations of NGC5044 were carried out during June 20--21 1993.
We used a representation of the surface brightness profile from ROSAT
PSPC data given by a $\beta$-model with parameters $\beta=0.51$,
$r_{a}=10.6$ kpc, obtained from our analysis of ROSAT PSPC data in the
0--40\amin\ region, excluding point sources and a ``cooling wake'', found
by David \etal\ (1994).  To compare with our ASCA results, we carried out
an identical 3-dimensional spectral analysis with ROSAT PSPC data.

\begin{figure}
\vspace*{-0.5cm}
\centerline{\scalebox{0.5}{\epsfbox{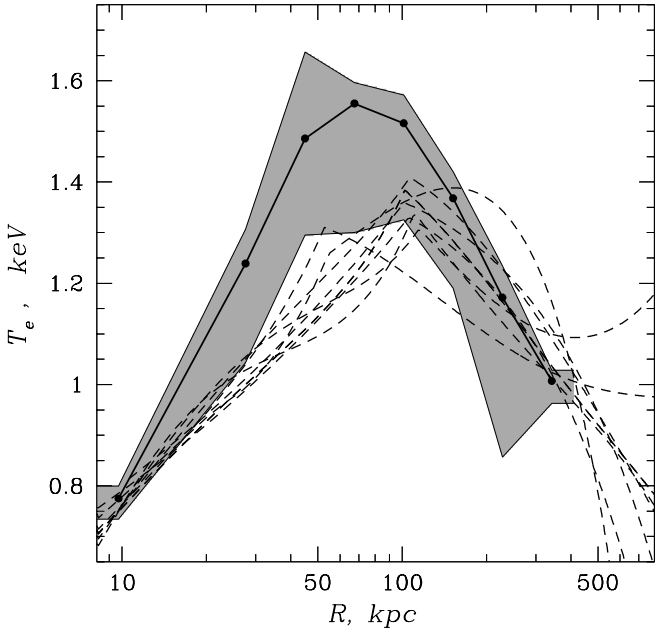}}}
\vspace*{-20pt} 
\caption{
Temperature profile of HCG62. Results of a three-dimensional analysis of
ROSAT PSPC data, fitting a variety of models to the data, are shown in
dashed lines. Filled circles indicating the spatial binning and the shaded
area showing the 90 per cent confidence level area around the best fit
(solid line) present the ASCA SIS results.}
\label{hcg62-te}
\vspace*{-0.2cm}
\end{figure}

In Fig.\ref{n5044te} we present a temperature profile of NGC5044. ROSAT PSPC
points are represented by black crosses, solid line represents the best-fit
curve describing ASCA results with filled circles indicating the spatial
binning used. Shaded zone around the best fit curve denotes 90 per cent
confidence area. The ASCA data confirm a pronounced cooling flow at the
centre with a temperature drop from 100~kpc to 10~kpc by a factor of two.
ASCA and ROSAT results agree at radii exceeding 30~kpc, with the
disagreement at small radii indicating the presence of complex temperature
structure in the central cooling flow. Our results agree well with the
previous finding of David \etal\ (1994) that between 60 and 250 kpc the gas
is nearly isothermal.

Element abundance profiles are shown in Fig.\ref{n5044ab}.  (We exclude the
cooling region, where abundance results may be unreliable.) A significant
decrease with radius is observed, and we see for the first time that this
applies to Mg and Si in a similar way to Fe.  The abundance of iron
determined from ROSAT data (grey points) also shows a decrease with radius.
The ASCA abundances drop from 0.3, 0.5 and 0.5 solar at $r=40$~kpc to near
0.0, 0.2 and 0.1 solar by $r=200$~kpc for Mg, Si and Fe, respectively. Shown
errors on Fe abundance are dominated by the assumed systematic uncertainty
in modelling.

\subsection{HCG62}

HCG62, is a compact group, taken from the catalogue of Hickson (1982),
which was compiled by locating compact configurations of galaxies on the
Palomar All Sky Survey in the E-band. HCG62 is the most luminous of the
Hickson groups in the X-ray (Ponman \etal, 1996) and an analysis of the
ROSAT PSPC data for HCG62 is presented in Ponman \& Bertram (1993). The ASCA
SIS observation analyzed here was carried out on January 14 1994.

\begin{figure}
\vspace*{-0.5cm}
\centerline{\scalebox{0.5}{\epsfbox{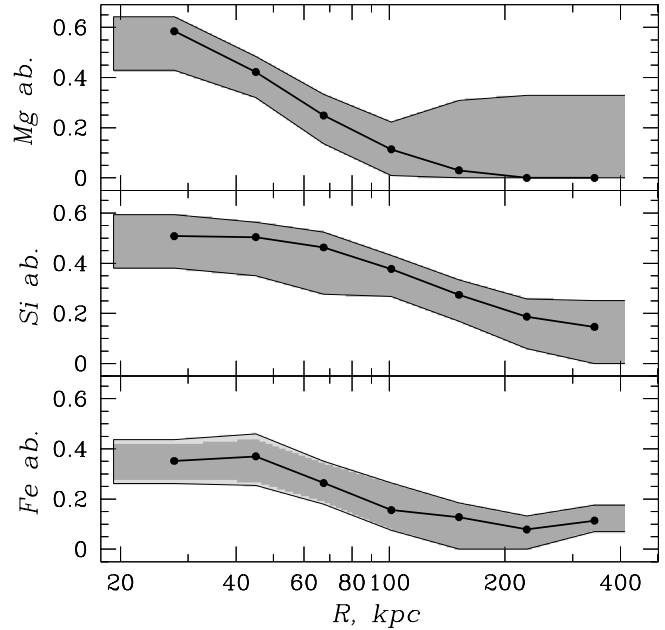}}}
\vspace*{-20pt} 
\caption{
 Mg, Si and Fe abundance profiles for HCG62 from ASCA data. Data
representation is similar to Fig.\ref{n5044ab}.}
\label{hcg62-feab}
\vspace*{-0.2cm}
\end{figure}

For our analysis we adopt a distance of 82~Mpc, which corresponds to a scale
1\amin $=$ 23.9~kpc. The X-ray surface brightness profile is characterized
by a double $\beta$-model fitted to the ROSAT PSPC data in the $r=0-20$
\amin\ interval.  The derived parameters of ($\beta$, $r_{a}$) are (0.66,
10.3 kpc) and (0.30, 52 kpc). Results of our analysis of ASCA SIS data were
then compared with the results of three-dimensional modelling of the ROSAT
PSPC data using the Birmingham cluster fitting package (Eyles \etal 1991),
which also allows for the effects of projection and energy dependent PSF
blurring, but adopts simple analytical models to represent the continuous
gas density and temperature distributions, rather than the regularised
discrete distributions employed in our ASCA analysis.

In Fig.\ref{hcg62-te} we compare the temperature distributions derived from
the ASCA and ROSAT data using these two different analysis systems.  In the
case of the ROSAT data, a variety of different models for the temperature
distribution have been fitted to the data, to give an indication of the
model dependence of the results. In general the two sets of results show an
encouraging level of agreement. The temperature profile is characterized by
cooling at the centre and a gradual decrease toward the edge of the group.

The element distributions derived from the ASCA analysis, presented in
Fig.\ref{hcg62-feab}, show strong gradients of Mg, Si and Fe abundance, in
the same sense as NGC5044. At $r=20$~kpc Mg and Si abundance are 0.4 solar,
Fe is 0.3 solar. All decline to $\sim$0.1 solar by a radius of 400 kpc. The
Fe abundance gradient implied by ROSAT data matches the ASCA data in the
outer parts, but rises to $\sim$0.6 solar at 40 kpc, although this
difference near the centre is still within the combined 90\% errors
from both analyses.

\subsection{HCG51}

\begin{figure}
\vspace*{-0.5cm}
\centerline{\scalebox{0.5}{\epsfbox{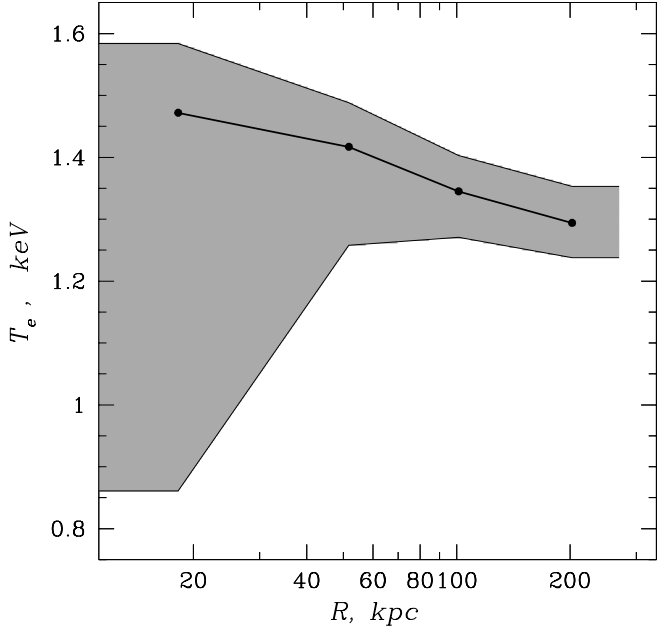}}}
\vspace*{-20pt} 
\caption{
Temperature profile of HCG51, derived from ASCA SIS data. The best-fit curve
does not indicate the presence of the cooling flow in the data, although the
uncertainty in the parameter estimation is large for the inner points.}
\label{hcg51-te}
\vspace*{-0.2cm}
\end{figure}

From results of the compact group survey of Ponman \etal\ (1996), HCG51 is
one of the most X-ray luminous of the compact groups. Apart from a short
observation in the ROSAT All Sky Survey, the group was never observed by the
ROSAT PSPC. However a 25~ksec observation was obtained in May 1995, and has
been reduced using the Starlink ASTERIX analysis system, to derive a surface
brightness profile for use in the ASCA data modelling procedure.

Diffuse emission is centered on the brightest elliptical galaxy in the
group, and is detected in the HRI data out to 6\amin\ from the group centre.
An adequate description of the profile in terms of $\beta$-models requires
the introduction of a second component giving parameters ($\beta$, $r_{a}$)
equal to (11.1, 33.2 kpc) and (0.30, 81 kpc).

ASCA observations of HCG51 were carried out on June 3--5 1994 using two-CCD
mode for the SIS. We adopt a distance of 154.7~Mpc to HCG51, which gives a
scale 1\amin$=$45~kpc.

\begin{figure}
\vspace*{-0.5cm}
\centerline{\scalebox{0.5}{\epsfbox{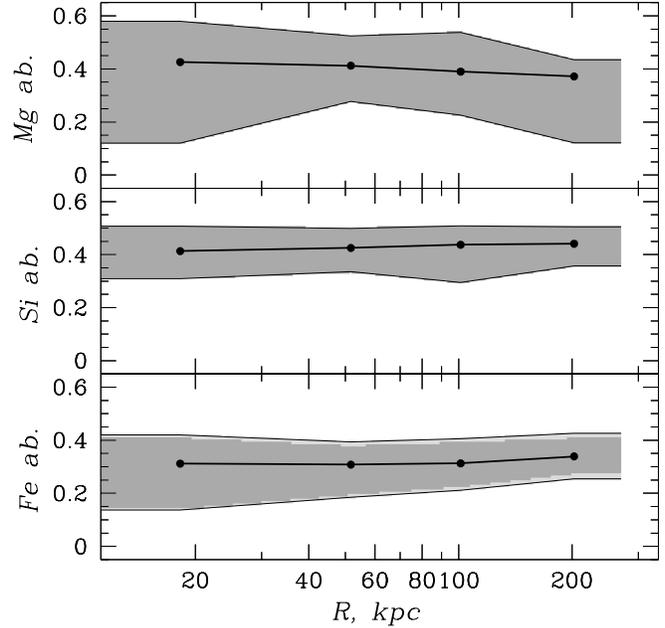}}}
\vspace*{-20pt} 
\caption{
Mg, Si and Fe abundance profiles in HCG51 derived from ASCA SIS data (data
representation is similar to Fig.\ref{n5044ab}). No significant gradients
are seen.}
\label{hcg51-ab}
\vspace*{-0.2cm}
\end{figure}

The derived temperature and abundance profiles are presented in
Fig.\ref{hcg51-te} and \ref{hcg51-ab}. In contrast to the other systems
studied here, HCG51 does not show pronounced signs of a cooling flow. The
temperature is $\sim1.35$ keV, with some indication of a gentle decline with
radius. Nevertheless, considering the confidence area for temperature
estimation, it is possible to fit in the central cooling zone. The derived
abundances of Mg, Si and Fe do not show any significant variation with
radius. The lack of pronounced central cooling and the flat abundance
profiles (contrasting with strong gradients in the other systems) strongly
suggest that recent gas mixing has taken place in HCG51.

\subsection{AWM7}

ASCA SIS observations of AWM7 were obtained during
August 7--8 1993 and February 10--12 1994. We adopt a distance to AWM7 of
105.6~Mpc, so that 1\amin$=$31~kpc. A surface brightness model was taken from
the work of Neumann \& Boehringer (1995), giving values of the parameters
($\beta$, $r_{a}$) in the two $\beta$-model approximation of (0.25, 5~kpc)
and (0.53, 102~kpc). The relative normalization of the components was chosen
to give a central density ratio of 2, as obtained by Neumann \& Boehringer
(1995).

\begin{figure}
\vspace*{-0.5cm}
\centerline{\scalebox{0.5}{\epsfbox{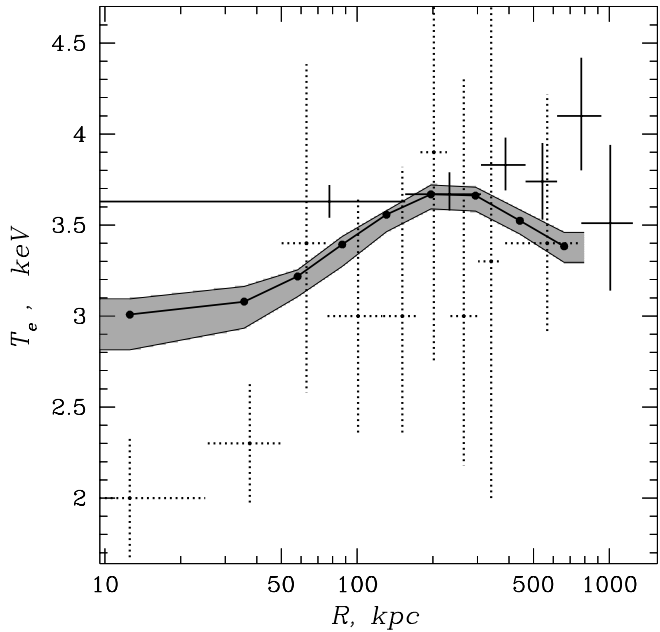}}}
\vspace*{-20pt} 
\caption{
Temperature profile of AWM7. Filled circles represent best fit to our ASCA
SIS data with shaded area showing the 90 per cent confidence level. Crosses,
denoted with a dotted line represent ROSAT/PSPC data from Neumann \& Boehringer
(1995), solid black crosses denote the results of Ezawa \etal (1997). All
confidence intervals on the plot are given at 90 per cent confidence.}
\label{awm7-te}
\vspace*{-0.2cm}
\end{figure}

In Fig.\ref{awm7-te} we compare the temperature profile derived here from
ASCA data with the previous findings of Neumann \& Boehringer (1995) and
also from an annular spectral analysis (neglecting projection effects) of
ASCA data by Ezawa \etal (1997). The latter authors have extended their
analysis to larger radii than the present work, due to a more detailed
treatment of stray light effects, but did not apply any regularization,
resulting in rather coarse spatial resolution. The three analyses are in
broad agreement apart from near the centre, where ROSAT data show the
strongest cooling flow, while the analysis of Ezawa \etal (1997) does not
reveal any temperature decrease at all. Our temperature profile is
intermediate, but disagrees significantly with the ROSAT results only
at $r<50$~kpc, indicating the probable presence of a multi-phase cooling
flow, as discussed by Markevitch \& Vikhlinin (1997).

\begin{figure}
\vspace*{-0.5cm}
\centerline{\scalebox{0.5}{\epsfbox{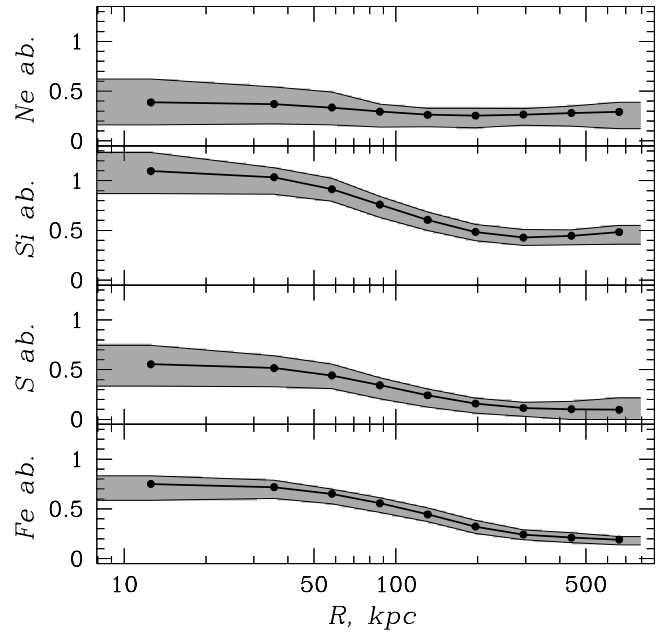}}}
\vspace*{-20pt} 
\caption{
Abundances of Ne, Si, S and Fe in AWM7, obtained from ASCA SIS data (data
representation is similar to Fig.\ref{n5044ab}).  An initial abundance
decrease is followed by flattening at a radius of $\sim$200~kpc.}
\label{awm7-ab}
\vspace*{-0.2cm}
\end{figure}

In Fig.\ref{awm7-ab} we present abundance profiles derived for Ne, Si, S and
Fe. All abundances show a decrease with radius which, with the possible
exception of Fe, flattens off outside 200~kpc. Central values are 0.4, 1.1,
0.6 and 0.7 solar, respectively, dropping at $r=$800~kpc to 0.3, 0.5, 0.1
and 0.2 solar. The strong cooling flow may be expected to affect derived
abundances in the inner regions, though the effects will be less severe than
in cooler systems. To gauge these effects we have fitted spectra from the
central 3\amin\ with a model including a cooling flow component. The
inclusion of such a component has a negligible effect on the derived Fe
abundance (which is largely based on the Fe K line for this system), whilst
abundances of Si and S rise by $\sim 35$ per cent, and Ne is poorly
constrained. Hence allowing for the effects of central cooling is only
likely to increase abundance gradients, and amplify the contribution of
$\alpha$-elements in the centre of the cluster, which we discuss below.

To investigate the systematic effects of ASCA PSF calibration
uncertainties, we explored the effects of varying the spatial binning
of the data. The conclusion we derived from these exercises is 
that the detected
abundance gradients are robust, exception for the central excess of sulphur
in AWM7, which was found to be  lower with coarser binning.

\section{Discussion}

Using the above results for the abundance distributions of iron and
$\alpha$-elements in these systems, we now proceed to examine the
contributions from SNe~II and SNe~Ia to the metal enrichment. We also
derive the iron mass to light ratio (IMLR) for
the intracluster gas in each system, which gives
clues about overall metal production and loss, and given our derived
supernova rates we evaluate the importance of energy injection by SNe.
Finally, we consider our results in the context of hierarchical
clustering models.

\subsection{Roles of different types of SNe in chemical enrichment of the
IGM}

The heavy elements liberated in type Ia and type II supernovae differ
markedly in the ratio of iron to the elements (O, Ne, Mg, Si and S)
generated by the alpha-process. For example, the Si/Fe abundance ratio is
at least 4 times higher in the ejecta of SN~II, compared to SN~Ia.
Unfortunately, the complexity of the theory of SN~II explosions leads to
a substantial uncertainty in SN~II yields between different models. First
attempts to confront the abundance measurements in clusters with
theoretical predictions were made by Loewenstein \& Mushotzky (1996).
Their conclusion was that for some models one doesn't need any SN~Ia to
reproduce the observed element ratios. They also
found that none of the models they considered could self-consistently
explain all the element ratios observed.

\begin{figure}
\vspace*{-0.5cm}
\centerline{\scalebox{0.5}{\epsfbox{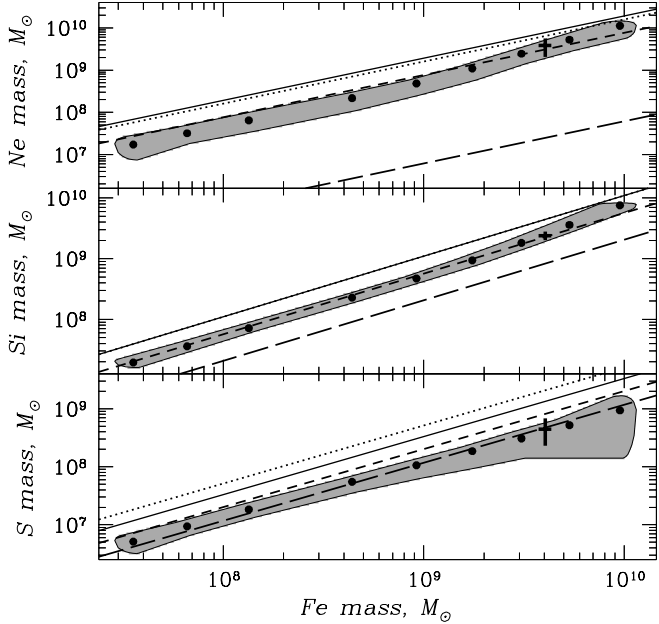}}}
\vspace*{-20pt} 
\caption{
Ne, Si and S mass vs Fe mass in AWM7. Crosses represent ASCA SIS data from
Mu96. Filled circles indicate our best-fit solution with shaded regions
denoting the 90 per cent confidence intervals.  
The long-dashed line shows the theoretical mass
ratio for SN~Ia yield (TNH93 model), whilst SN~II yields from the
compilation of Gibson \etal (1997) are shown by solid and dotted lines
denoting a choice of T95 and W95 yields, respectively. (These coincide 
in the case of Si.) The dashed line
represents a 60 per cent contribution of SN~Ia to the iron enrichment, using
T95 yields for SN~II.}
\label{awm7rat}
\vspace*{-0.2cm}
\end{figure}

Turning to our observation of AWM7 -- this is one of the objects studied by
Loewenstein \& Mushotzky (1996, LM96), based on the measurements of
Mushotzky \etal (1996, Mu96), but we are able to build a more detailed
picture, since we have resolved the radial element distributions. SN~II
yields are taken from the recent work of Gibson \etal (1997), where an
attempt was made to put different theoretical predictions onto the same mass
grid for more direct comparison. We use models T95 and W95(A,Z$_{\odot}$),
from Gibson \etal (1997), which are both integrated over a Salpeter IMF. SN
Ia yields are subject to much less uncertainty, and are taken from
Thielemann \etal (1993, TNH93). Average stellar yields for Ne, Mg, Si, S,
Fe in solar masses are 0.232, 0.118, 0.133, 0.040, 0.121 for the T95 model;
0.181, 0.065, 0.124, 0.058, 0.113 for the W95 
model; 0.005, 0.009, 0.158, 0.086, 0.744 for the TNH93 model.

\begin{figure}
\vspace*{-0.1cm}
\begin{minipage}{4.cm}
\centerline{\scalebox{0.25}{\epsfbox{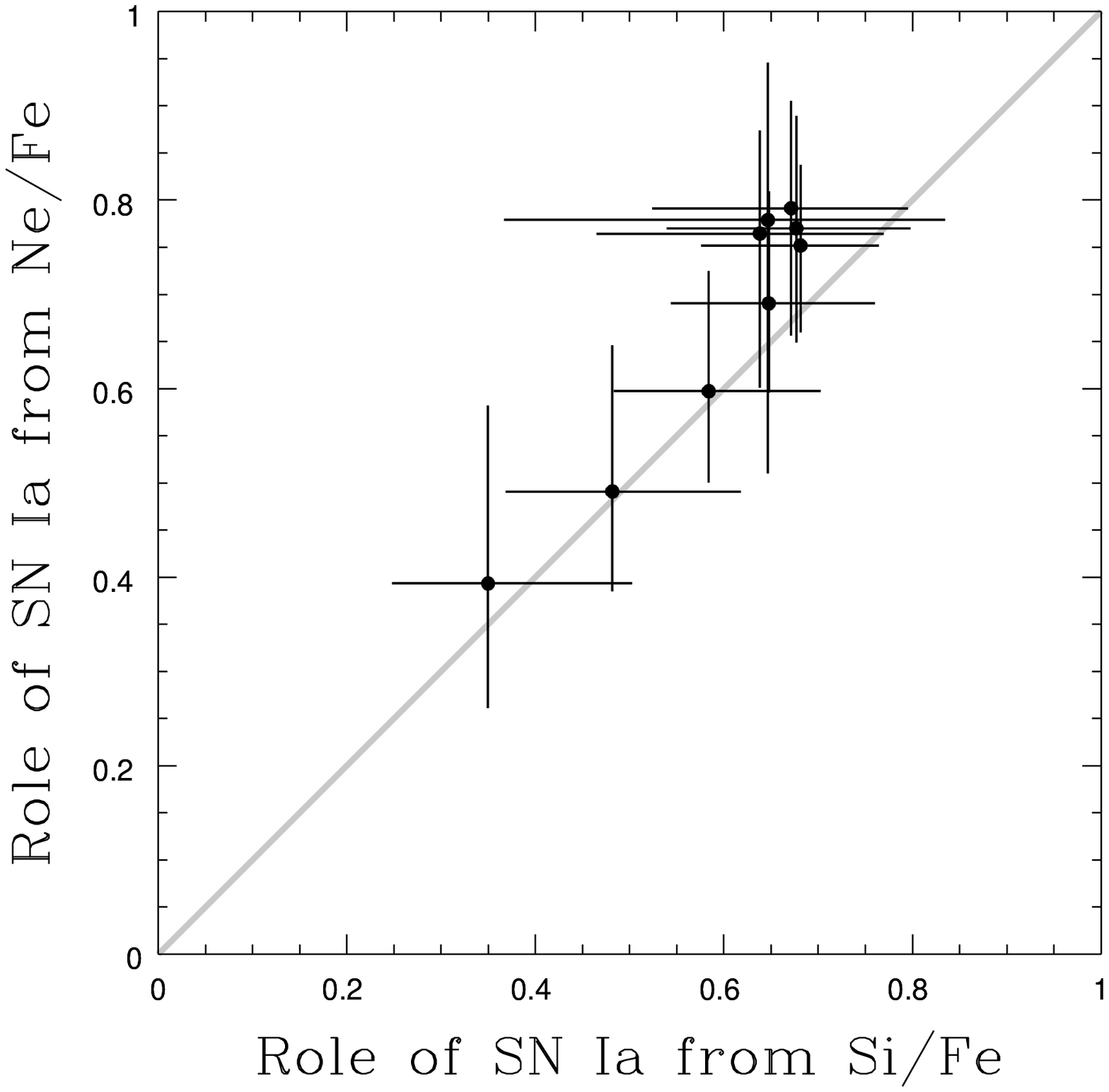}}}

\end{minipage} \hspace{0.1cm} \begin{minipage}{4.cm}
\centerline{\scalebox{0.25}{\epsfbox{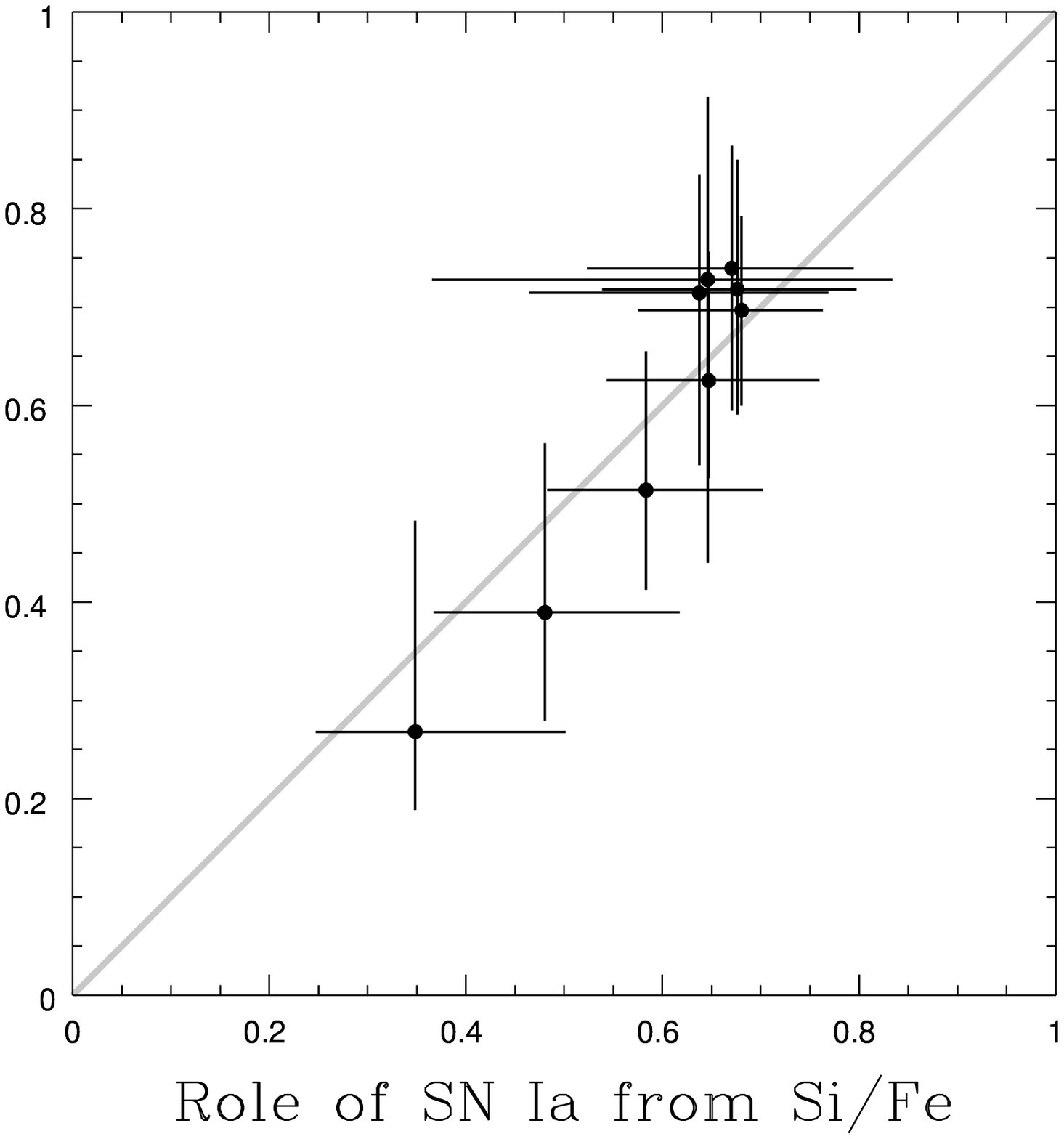}}}

\end{minipage}
\vspace*{-10pt} 
\caption{
Input of SN~Ia into the iron mass, calculated from Si/Fe and Ne/Fe ratios,
assuming T95 ({\it left panel }) and W95 ({\it right panel}) yields for
SN~II. Points correspond to measurements at different radii in AWM7. Error
bars are 1$\sigma$, and the points are not independent due to
regularization.}
\label{awm7-ne-si}
\vspace*{-0.2cm}
\end{figure}

In Fig.\ref{awm7rat} we consider the behaviour of the cumulative masses of
$\alpha$-elements plotted against the cumulative mass of iron. 
The shaded zone represents the 90 per cent confidence area for
deviation from either parameter, so the form of the shaded area at the
boundary points has an elliptical form. The element ratios predicted for
SN~Ia and the chosen models of SN~II are shown by straight lines. In
addition we plot a line corresponding to a 60 per cent contribution of SN~Ia
to the iron mass, calculated using the T95 model for SN~II yields and the
TNH93 model for SN~Ia yields. Also marked are the results of Mu96, which
agree with our values well at comparable radii. Note that Si/Fe (and Ne/Fe)
is lower than the Mu96 value at smaller radii, but continues to rise towards
SN~II-like ratios at larger radii.

An anomaly immediately apparent in Fig.\ref{awm7rat}, is that whilst Ne and
Si masses show similar behaviour in the SN~Ia -- SN~II reference frame, the
S/Fe ratio is approximately constant through the cluster. The anomalous
behaviour of the S/Fe ratio was also noted in integrated spectra by Mu96. We
note that not only the absolute value, but also the trend in the S abundance
is anomalous. The observed S/Fe ratio is consistent with 100 per cent Fe
production by SNe~Ia throughout the cluster, while from other elements it
varies from 70 per cent down to 30 per cent, from inner to outer radial
points. Such behaviour appears to imply that the SN~II model yields for
sulphur are incorrect or that the observed abundance is modified by some
additional physical process.

Leaving sulphur aside as unreliable, we can derive independent estimates of
the fraction of the iron which is contributed by SN~Ia from the Si/Fe and
Ne/Fe ratios, using a given pair of models for the SN~Ia and SN~II yields.
This can be used as a test of the consistency of the models. In
Fig.\ref{awm7-ne-si} we compare the results for AWM7 using the T95 and W95
models for SN~II (the TNH93 model yields are adopted for SN~Ia in each
case).  Both models appear acceptable in this case. For comparison, the
SN~II model of Thielemann \etal (1996; yields for Fe and Si are 0.11 and
0.08 in solar masses), which LM96 note can give a reasonable match to most
of their integrated abundances without any SN~Ia contribution, fails to
explain the high Si/Fe ratio which we observe at the outer radii in our AWM7
analysis.

\begin{figure*}
\vspace*{-0.5cm}
\begin{minipage}{5.2cm}
\centerline{\scalebox{0.3}{\epsfbox{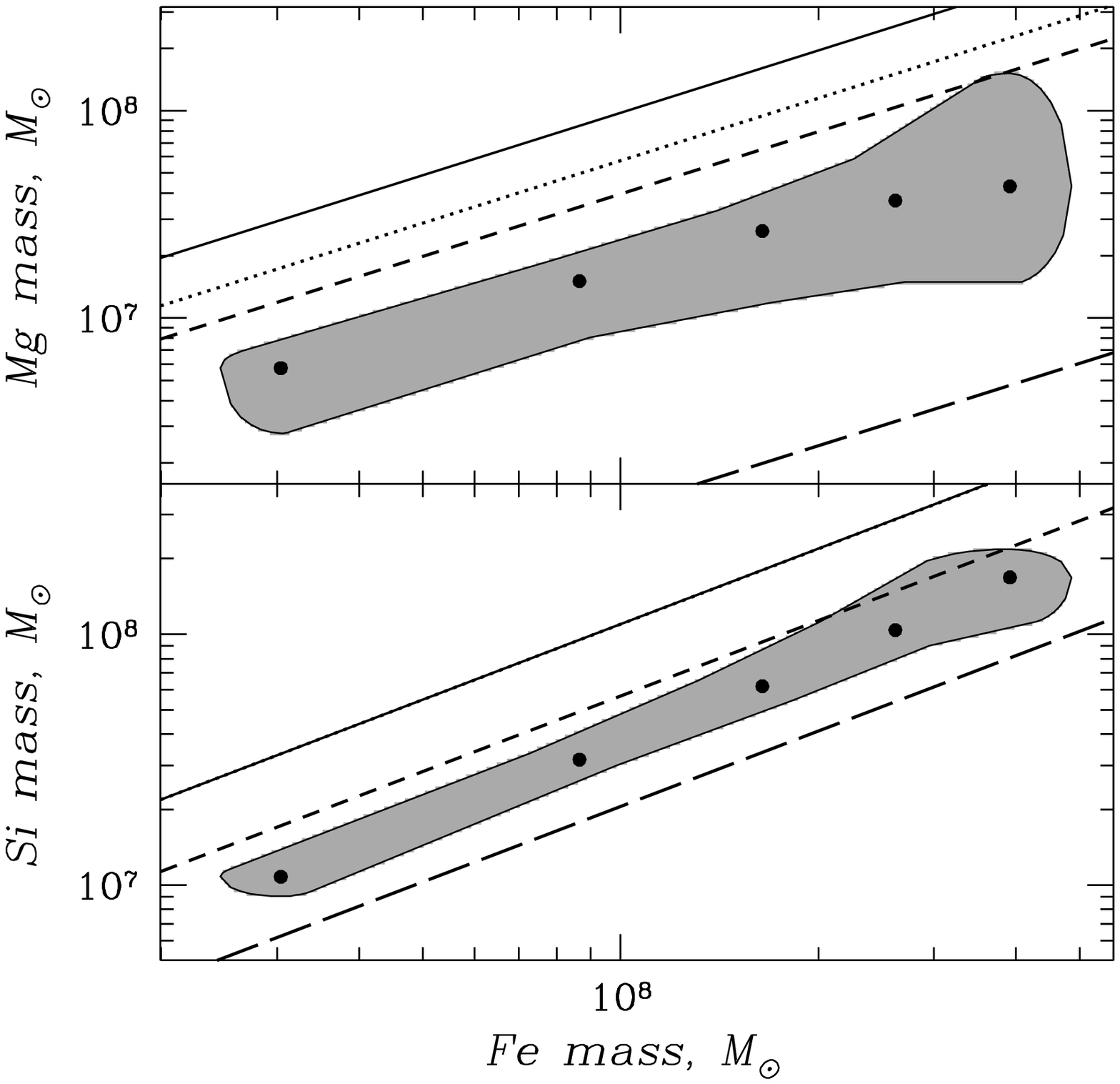}}}
\end{minipage} \hspace{0.45cm} \begin{minipage}{5.2cm}
\centerline{\scalebox{0.3}{\epsfbox{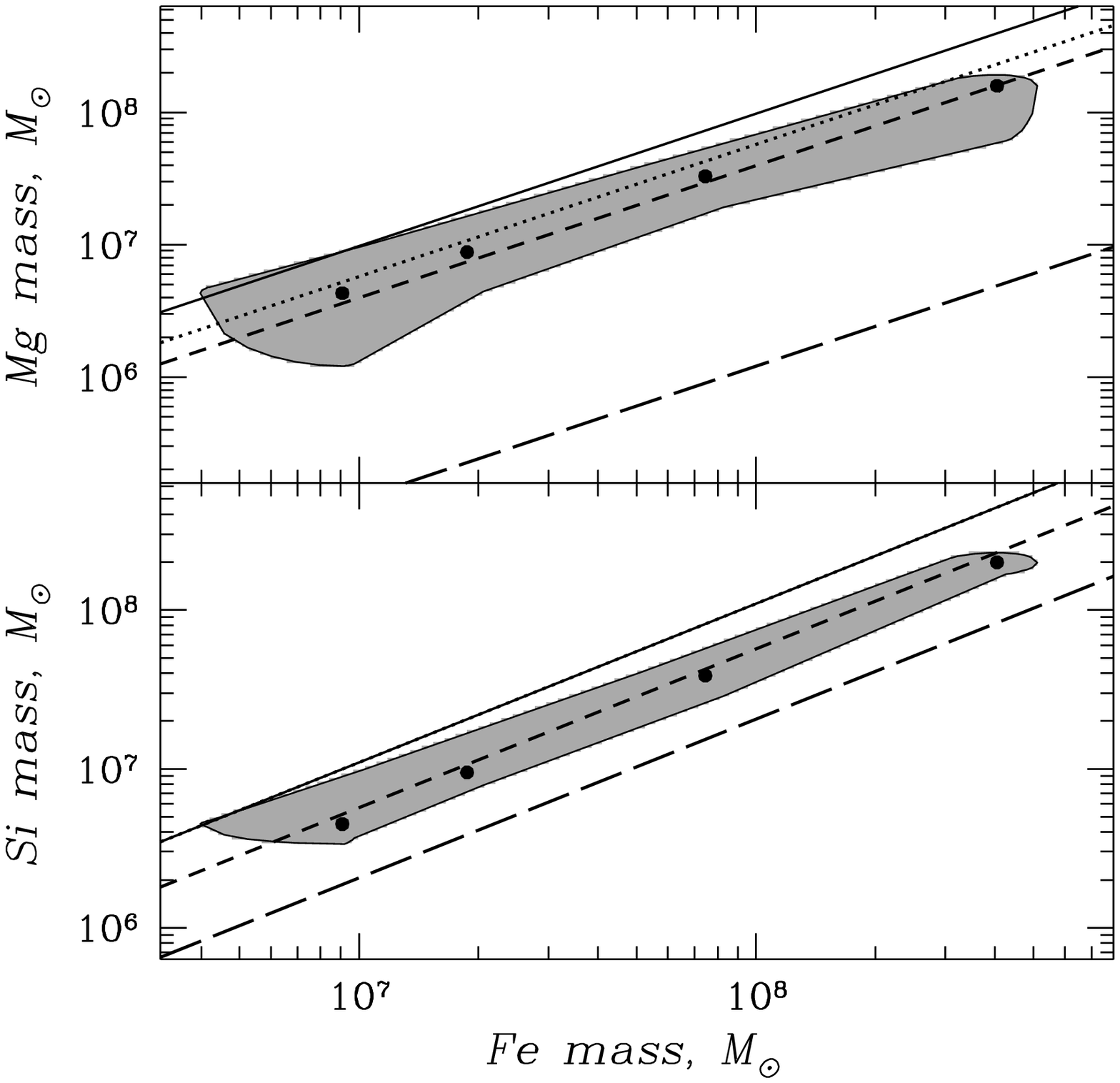}}}
\end{minipage} \hspace{0.45cm} \begin{minipage}{5.2cm}
\centerline{\scalebox{0.3}{\epsfbox{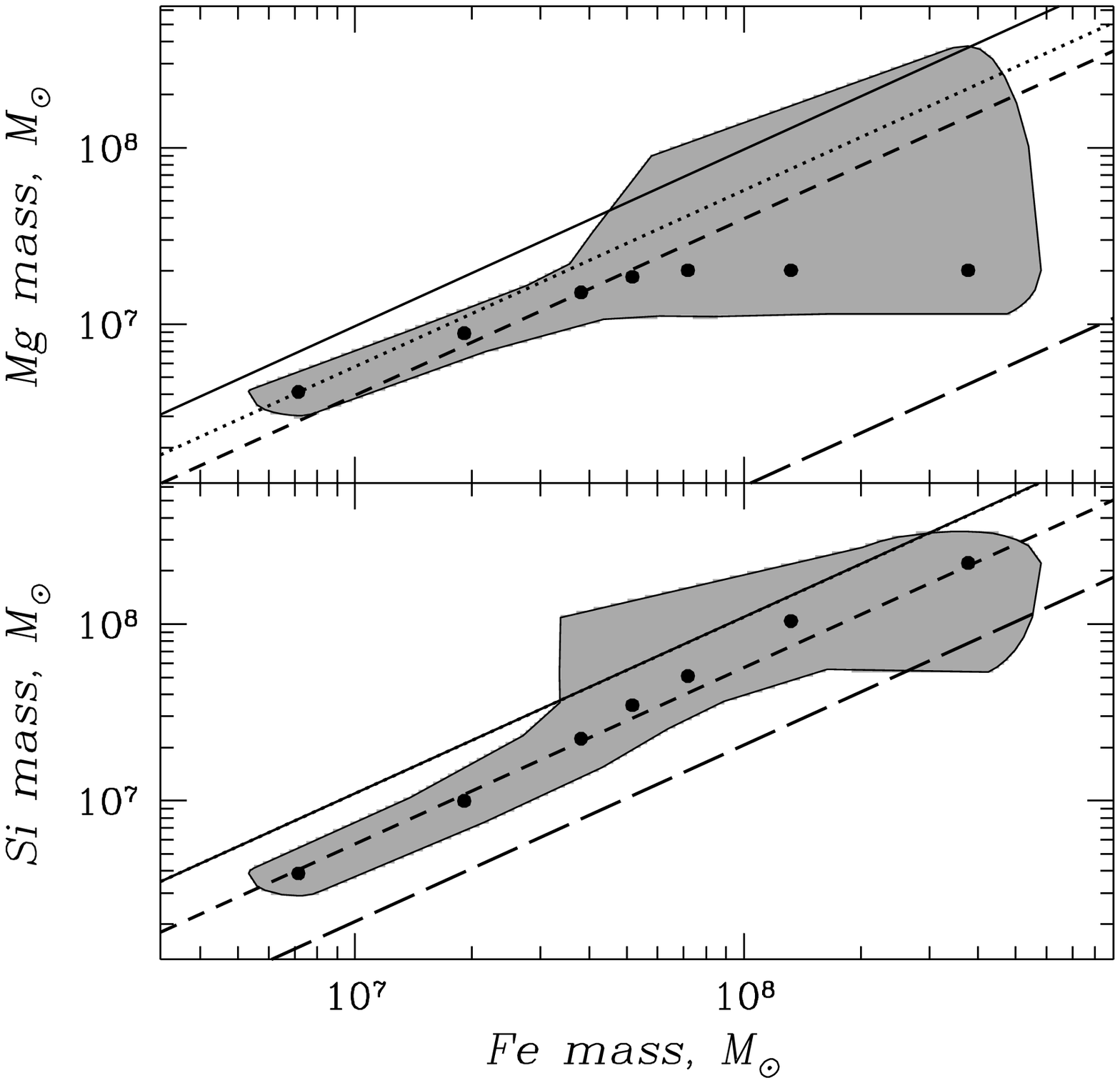}}}

\end{minipage}
\vspace*{-15pt} 
\caption{
Mg and Si mass vs Fe mass in NGC5044 ({\it left}), HCG51 ({\it center}) and
HCG62 ({\it right}). Line coding for the theoretical SNe yields and data
representation is similar to Fig.\ref{awm7rat}.}
\label{grp-rat}
\vspace*{-0.2cm}
\end{figure*}

\begin{figure}
\vspace*{-0.1cm}
\begin{minipage}{4.cm}
\centerline{\scalebox{0.25}{\epsfbox{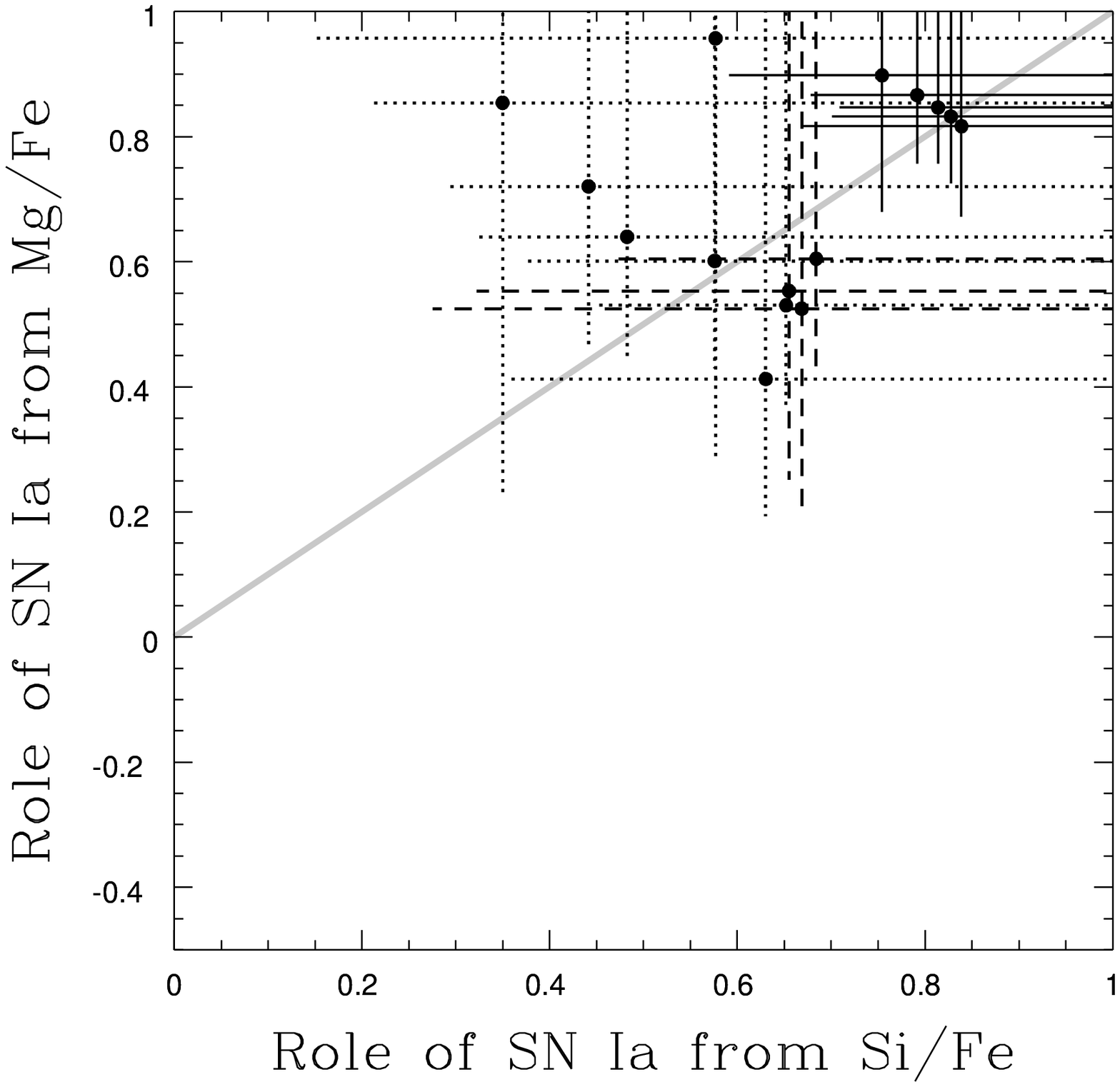}}}

\end{minipage} \hspace{0.1cm} \begin{minipage}{4.cm}
\centerline{\scalebox{0.25}{\epsfbox{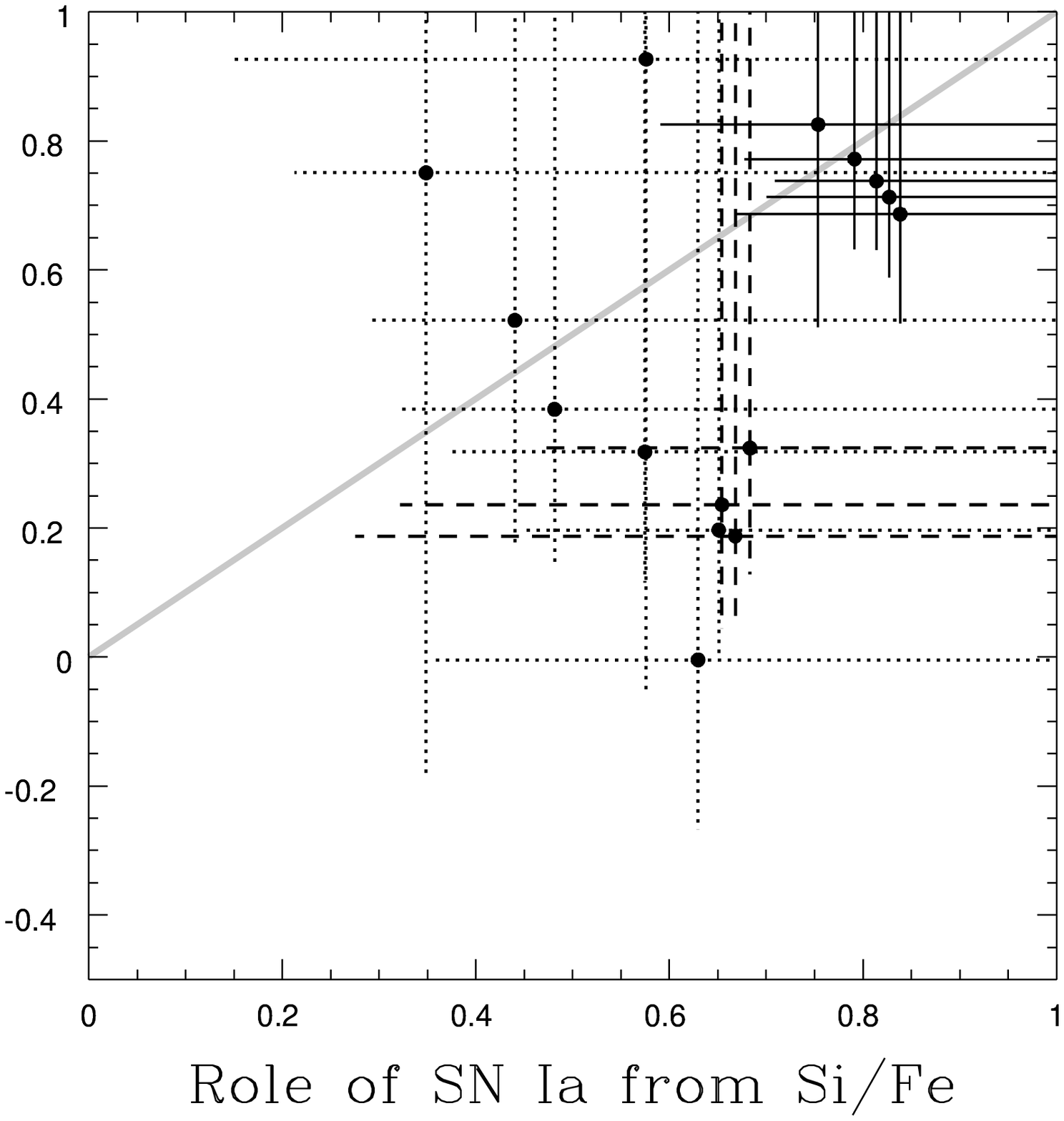}}}

\end{minipage}
\vspace*{-10pt} 
\caption{
  Input of SN~Ia into the iron mass, calculated from Si/Fe and Mg/Fe ratios,
  assuming T95 ({\it left panel}) and W95 ({\it right panel}) yields for
  SN~II. Solid line denotes NGC5044 data, dashed -- HCG51 and dotted --
  HCG62. Error bars are 1$\sigma$, and the points are not independent due to
  regularization.}
\label{all-mg-si}
\vspace*{-0.2cm}
\end{figure}

Our data on the abundance ratios in the groups are presented in
Fig.\ref{grp-rat}. Again, these ratios indicate significant contributions
from both SN~Ia and SN~II, but the variation of the element ratios with
radius is less pronounced than in AWM7. The role of SNe~Ia in iron
production is the strongest in NGC5044, where they provide $75\; (54-98)$ per
cent of the iron within the region analyzed, while in HCG62 and HCG51 the
corresponding values are somewhat lower and amount to $48\; (10-100)$ and $68\;
(40-95)$ per cent, respectively. Consistency between the estimates of the
SN~Ia contribution from Si/Fe and Mg/Fe is somewhat better for the T95 model
than for W95. But both models are acceptable within the errors, as is
illustrated in Fig.\ref{all-mg-si}.

The need for a significant SN~Ia contribution to explain the
abundance ratios in the systems we study has interesting implications
for the debate on the chemical evolution of elliptical galaxies.
In particular, one of the options discussed by Arimoto \etal (1996)
to account for the apparently low abundances seen in the hot
gas within ellipticals, was that the SN~Ia rate has always been
low. This option is ruled out by our results, and also by the findings
of Fukazawa \etal (1998).

\subsection{Iron profiles for SN~Ia and SN~II }

The iron abundance gradients seen in most of our systems show that the IGM
is not completely mixed. In addition to Fe, abundance gradients are detected
for $\alpha$-elements: Mg and Si in HCG62 and NGC5044, and Ne, Si and S in
AWM7. As discussed above in relation to AWM7, the balance between the
contribution of different SNe types could be a variable function of radius.
To explore this further, we use the ratio of $\alpha$-elements to iron as a
function of radius to decompose the iron profile into radially varying
contributions from SN~Ia and SN~II. The situation is complicated by the
likelihood that there will also be some contamination of the IGM by metals
released within galaxies by stellar mass loss.  Assuming that most of the
metals in the IGM are liberated from early-type galaxies (Arnaud \etal\
1992) the abundance ratios of such winds would be similar to that of SN~II,
so this contribution will effectively be included in the SN~II component in
our decomposition.  As we will see in the next section, the stellar wind
contribution, if it can escape from galaxies, may be significant, but should
not be a dominant component in the observed mass of metals.

\begin{figure}
\vspace*{-0.5cm}
\centerline{\scalebox{0.5}{\epsfbox{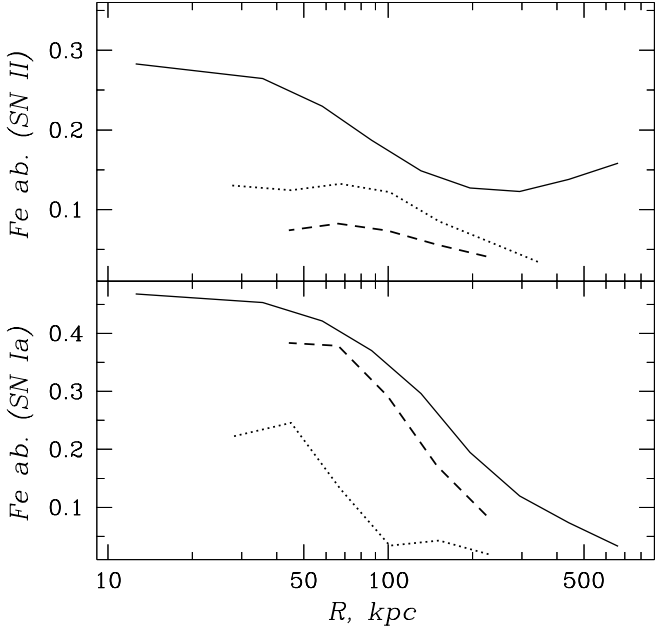}}}
\vspace*{-20pt} 
\caption{
Iron profiles decomposed into the two major types of SNe. Solid lines represent
AWM7 data, dashed line -- NGC5044, dotted -- HCG62. Results for HCG51 are
not shown since abundances show no significant variations.}
\label{iron_sne}
\vspace*{-0.2cm}
\end{figure}

For our decomposition of the iron profile into the two major SN types, we
use the Si and Fe abundance measurements and the yields of T95. This
approach has the advantages that Si and Fe are measured for all systems
studied here, the Si abundance is the best constrained of the 
$\alpha$-elements, and the spread in the Si/Fe yields between the T95 model
and the various models of W95 is small. This provides more confidence in the
derived results. In Fig.\ref{iron_sne} we present the iron profile
attributed to SN~II (upper panel) and SN~Ia (lower panel) for all each of
our systems apart from HCG51, which has essentially uniform abundances of
all elements, as was shown earlier.

The main feature seen in Fig.\ref{iron_sne} is the strong central
concentration of the SN~Ia iron profile, in comparison to a flatter
distribution in the SN~II iron contribution. Note that the fact that the
Si/Fe ratio varies significantly with radius in three of these systems
implies the need for at least two different sources of enrichment,
irrespective of the details of specific models for supernova yields.

To analyze the central excess in the SNe~Ia contribution, we fitted a King
profile. The values of the fitted core radii are 60 (27--121)~kpc for
HCG62, 117 (65--213)~kpc for NGC5044 and 171 (104--262)~kpc for AWM7, which
compares with their corresponding optical core radii of 27, 180 and 230~kpc.
We return to the broad similarity between these core radii below.

If it is assumed that SN~II ejecta escape primarily from early-type
galaxies, where they are generated by an initial massive burst of star
formation accompanying the birth of these galaxies (e.g. Matthews 1989), or
by subsequent bursts triggered by galaxy merging ( Schweizer \& Seitzer
1992), then their injection will be closely associated with active star
formation, and take place via highly energetic winds. The fact that the
SN~II products are so widely distributed, in contrast to the SN~Ia ejecta,
which are expected to be released over a much longer timescale after star
formation, suggests that the bulk of the SN~II activity occurred early in
the life of the system. This is consistent with the picture of Ponman,
Cannon \& Navarro (1999), who find that most of the supernova energy
injection must actually precede the collapse of groups and clusters in order
to account for the magnitude of the rise seen in the entropy of the gas.

The total level of Fe attributed to SN~II products
indicates the ability of the system to retain SN~II-driven winds.
Groups are characterized by a small spread around a value of 0.1 solar,
with average values of $0.07\pm0.02$ for NGC5044 and $0.11\pm0.03$ for
HCG62, however the SN~II contribution in both systems drops below
0.05~solar at large radii. The SN~II contribution to the well mixed IGM
in HCG51 is equal to $0.10\pm0.02$, within the region analyzed.

In AWM7, the SN~II contribution to the Fe abundance outside the core is
reasonably flat, at a level of $0.13\pm0.01$ solar. In the core of the
cluster, the SN~II contribution rises by a factor of two. This central
excess, which can be characterized by a King profile with a 55
(30--90)~kpc core radius, has some similarity with ASCA results for A1060
(Tamura \etal 1996), where a concentration of $\alpha$-elements towards
the cluster centre was detected. Both clusters are nearby and it is quite
probable that we are resolving the contribution to the metals arising
from stellar wind losses in the central cD galaxy. The fact that the IMLR
in the central region of AWM7 is below the level expected from stellar
mass loss (see next section) supports this suggestion.  However, in
NGC5044, which also contains a dominant central galaxy, no such excess
exists beyond the inner radius of 40~kpc employed in our analysis.

\subsection{Iron mass to light ratios}

\begin{figure}
\vspace*{-0.5cm}
\centerline{\scalebox{0.5}{\epsfbox{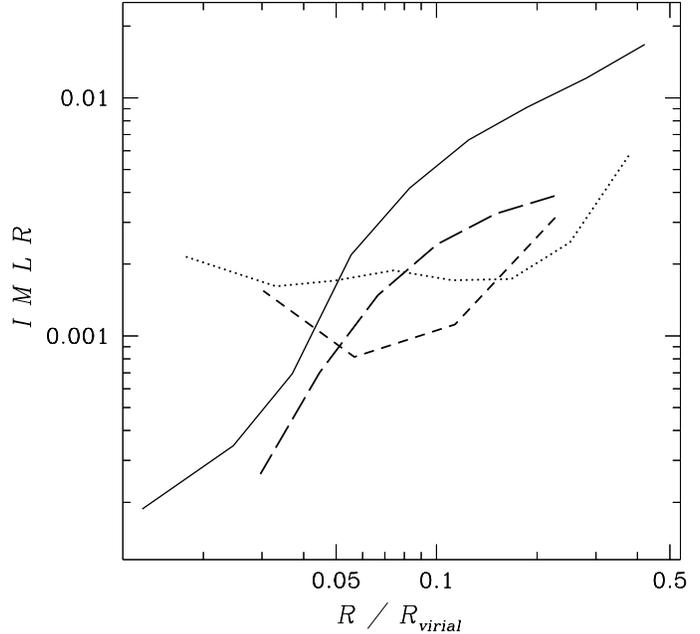}}}
\vspace*{-20pt} 
\caption{
Spatially resolved IMLR. The solid line represents AWM7 data, long dashed
line -- NGC5044, short-dashed -- HCG51, dotted -- HCG62. Errors do not
exceed 20 per cent and are negligible compared to
trends seen in the Figure.}
\label{all-imlr}
\vspace*{-0.2cm}
\end{figure}

\begin{figure*}

\begin{minipage}{8.cm}
\centerline{\scalebox{0.5}{\epsfbox{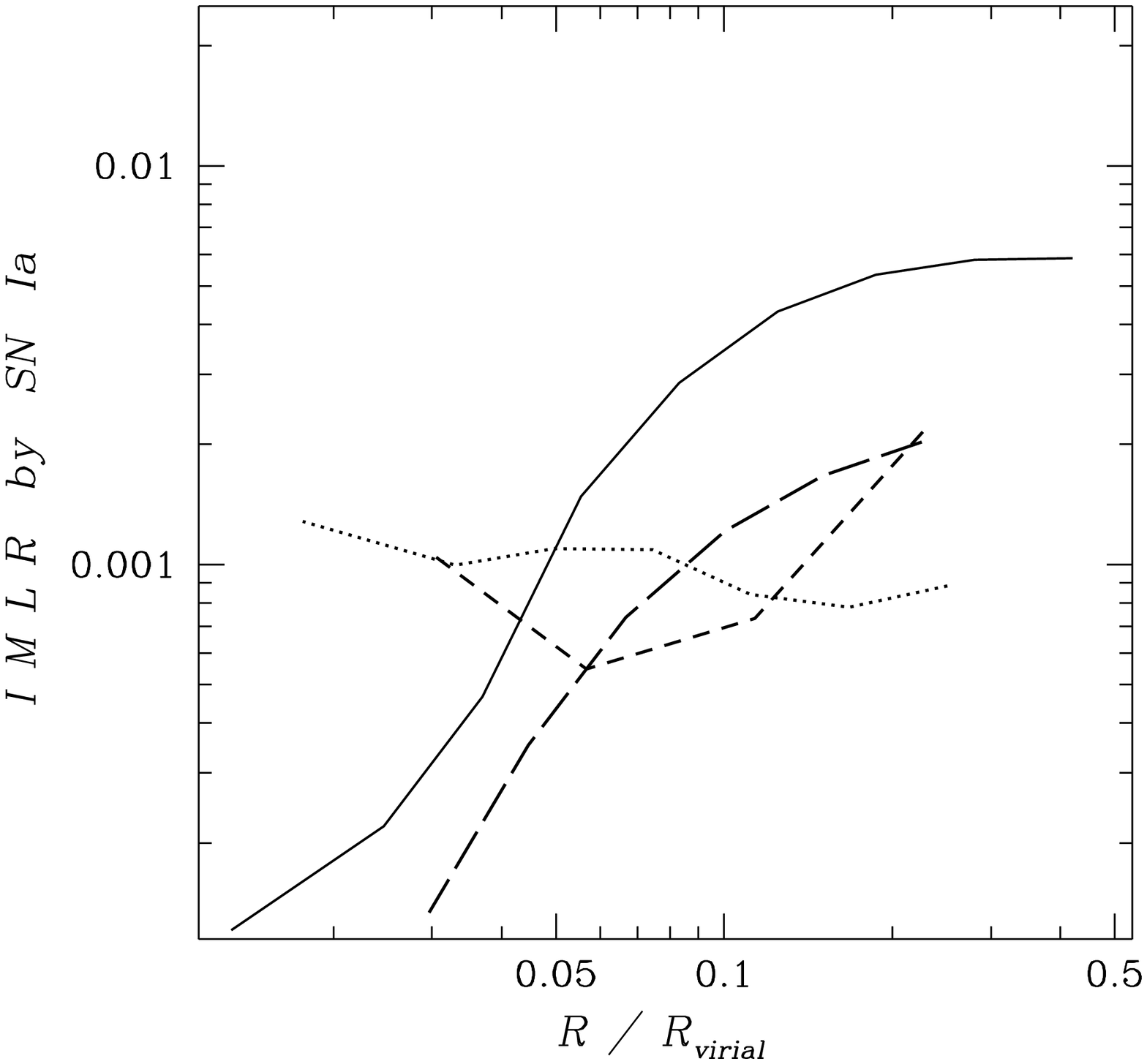}}}

\end{minipage} \hspace{1.cm} \begin{minipage}{8.cm}
\centerline{\scalebox{0.5}{\epsfbox{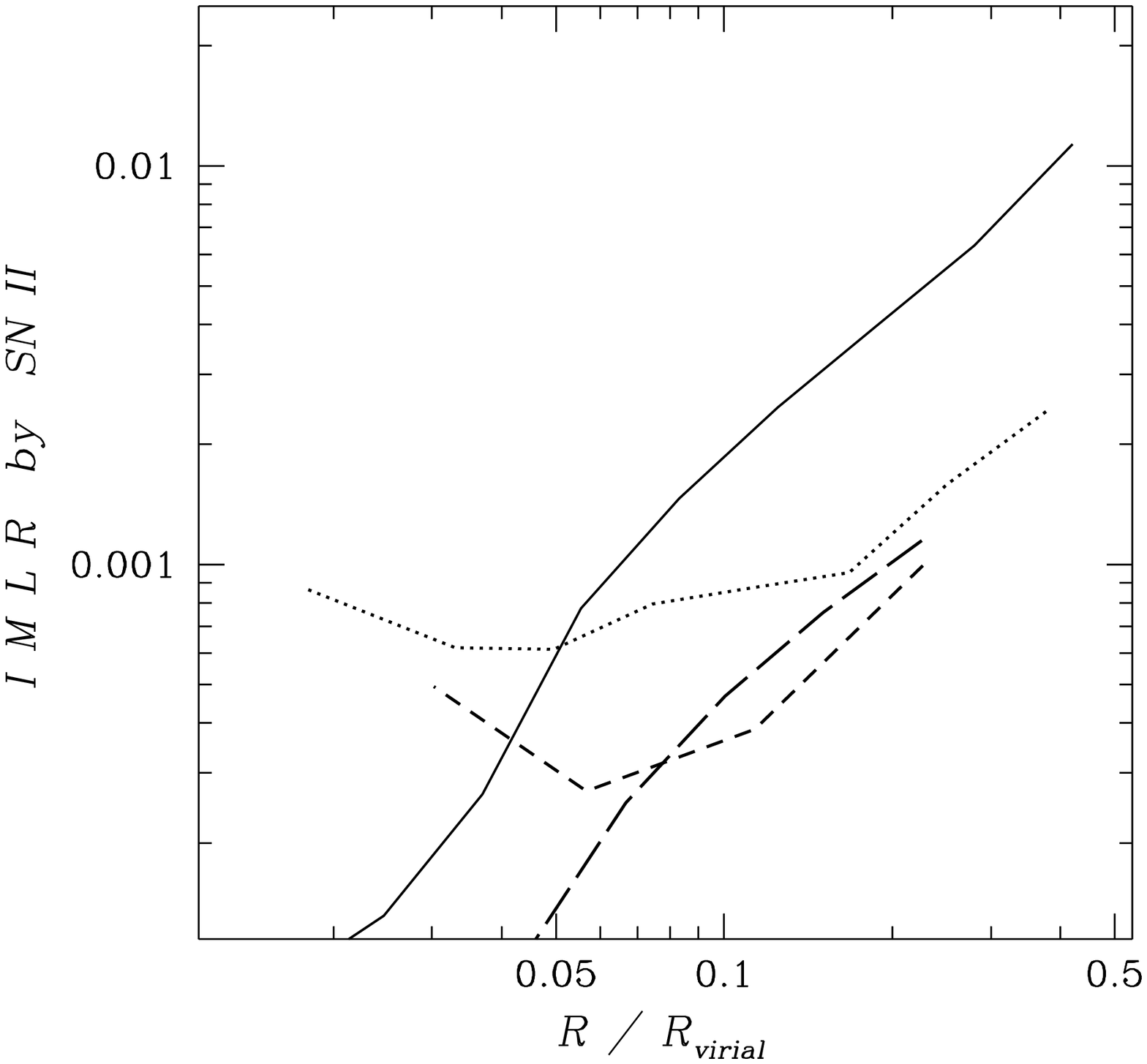}}}

\end{minipage}
\vspace*{-20pt} 
\caption{
Spatially resolved IMLR decomposed on types of SN: Ia {\it (left panel)} and II
{\it (right panel)}. Line coding is similar to Fig.\ref{all-imlr}.}
\label{all-sn-imlr}
\vspace*{-0.2cm}
\end{figure*}

The IMLR is a comparison of the amount of heavy elements with optical
luminosity, as an indicator of the ability of a system to synthesize and
retain elements. Having resolved the spatial behaviour of the elements, it
is straightforward to compare it with the {\it distribution} of optical
light, rather than an integral value. In the following, we will assume that
the distribution of galaxies in our systems is adequately described by a
King profile, characterized by some core radius ($R_c$). In addition we
separate the central galaxy from the distribution in the cases of AWM7 and
NGC5044, where it contributes a substantial fraction of the total light.
This means effectively an addition of a central "point" of light.
Parameters characterizing the optical luminosity, $L_{B, cD}$, $L_B$,
$R_{L_B}$ and $R_{c}$, are presented in Table~\ref{tab:opt}.

We normalize the distribution to the observed values ($L_B$) within some
radius ($R_{L_B}$) for AWM7 and NGC5044. For Hickson groups we require the
observed blue luminosity (Hickson \etal\ 1992) to be reached at the virial
radius of each system.  According to the study of Arnaud \etal (1992) the
metal content in clusters is correlated with the total blue luminosity of
early-type member galaxies rather than with the late-types. The values of
$L_B$, listed in Table 1 and used in calculating the IMLR, therefore
includes only the contribution from early-types. In the case of compact
groups there is good evidence that the core galaxies cataloged
by Hickson are accompanied by a broader distribution of mostly fainter
galaxies. This may increase the optical luminosity somewhat. In the
case of HCG62, the study of De Carvalho \etal (1997) indicates that
the optical luminosity we use might be increased by up to one third, 
if all the extra galaxies (which have not been typed) were early-types.

A comparison of the integrated IMLR in the IGM, plotted as a function of
$r/R_{virial}$, is shown in Fig.\ref{all-imlr}. The IMLR profiles for the
three groups are similar at radii exceeding $0.05R_{virial}$. The IMLR
reached in AWM7 at the outer radius of our analysis is similar to the
integrated value of 0.02 which is typical of clusters (Arnaud \etal\
1992, Renzini \etal\ 1993), whilst at a given radius, the groups have an
IMLR which is a factor of $\sim$3 times lower.

At small radii, there is a similarity between AWM7 and NGC5044 on one hand
and HCG51 and HCG62 on the other, arising from the presence of cD galaxies
in the former pair. In all our systems IMLR increases with radius with no
evidence for convergence, however we can map the element distribution only
to one half of the virial radius, at best. This rise in IMLR results from
the fact that the gas fraction increases with radius (the distribution of
the gas is flatter than that of the galaxies) which outweighs the decline in
abundance with radius.

Theoretical estimates suggest that the IMLR provided by
mass loss from stars with solar abundance could reach 0.001 over a
Hubble time (if it could all escape from galaxies), assuming a mass to
light ratio of 8, in solar units, and a Salpeter IMF (values for a
stellar mass loss are taken from Mathews 1989).  Comparison with
Fig.\ref{all-imlr} shows that this could be a significant contributor to
the metallicity, but is likely to be {\it dominant} only in the centres
of AWM7 and NGC5044. As we argued earlier, any stellar wind contribution
will be included in our analysis primarily within the SN~II contribution.
The fact that this appears to be distributed quite differently from the
SN~Ia ejecta also supports the idea that it is dominated by direct SN~II
ejecta, rather than stellar wind material.

In Fig.\ref{all-sn-imlr} we show a comparison of the IMLR in the systems,
decomposed into SN type. While a difference between the groups and AWM7 in
retention of SN~II products might be expected, we observe that even SN~Ia
products appear to have been lost from galaxy groups, by comparison to AWM7.
At $0.2 R_{virial}$ the level of IMLR attributed to SN Ia is $0.0009\;
(0.0004-0.0020)$, $0.0021\; (0.0014-0.0028)$ and $0.0020\; (0.0015-0.0025)$
for HCG62, HCG51 and NGC5044, respectively, while for AWM7 it amounts to
$0.0053\; (0.0040-0.0067)$. The conclusion that the SNIa contribution is
lower in the groups still holds if the W95 yields for SNII are used instead
of those from T95.

As we noted previously, the SN~Ia [Fe/H] distribution is rather similar
to the distribution of the optical light. However, [Fe/H] is produced by
mixture of the released gas with the primordial gas content of the
system. Let us assume, for simplicity, that at the time of gas release,
gas density, galaxy and dark matter followed similar profiles. Then, if
the efficiency of gas release from galaxies was proportional to the gas
density of the system, the resulting abundance distribution would be
flat. Hence to produce the observed abundance profile, the efficiency of
the gas release should go approximately as the square of the density.
Possible mechanisms expected to show such strong density dependence
include gas ablation, dark halo stripping (allowing ISM to escape more
readily) and galaxy-galaxy interactions (triggering starburst activity
and hence galaxy winds). The apparent loss of SN~Ia products from groups
suggests that injection is an energetic process, which favours galaxy
winds (possibly facilitated by dark halo stripping) over ram pressure
stripping of gas.

\begin{figure*}
\vspace*{-0.5cm}
\begin{minipage}{8.cm}
\centerline{\scalebox{0.5}{\epsfbox{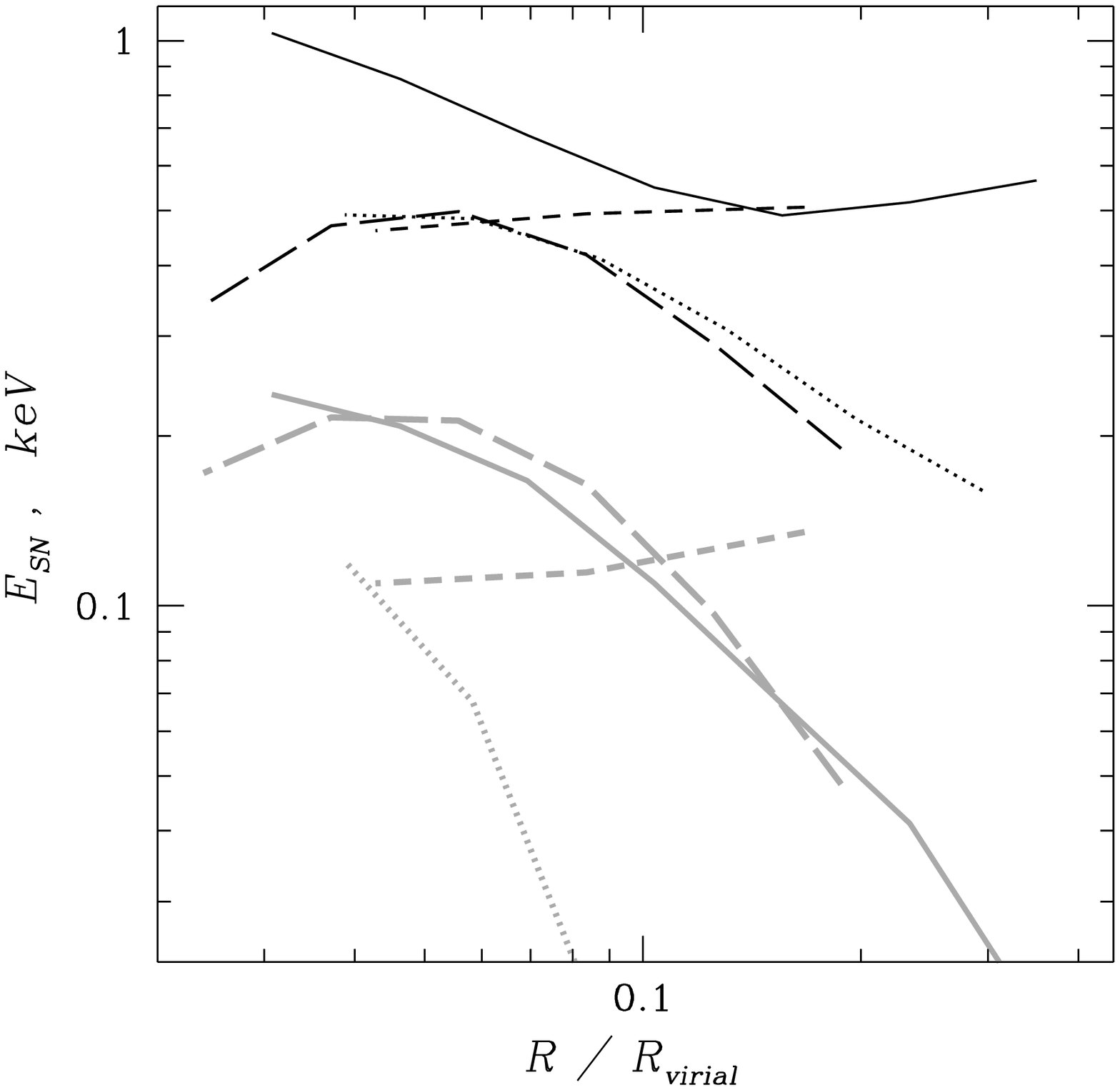}}}

\end{minipage} \hspace{1.cm} \begin{minipage}{8.cm}
\centerline{\scalebox{0.5}{\epsfbox{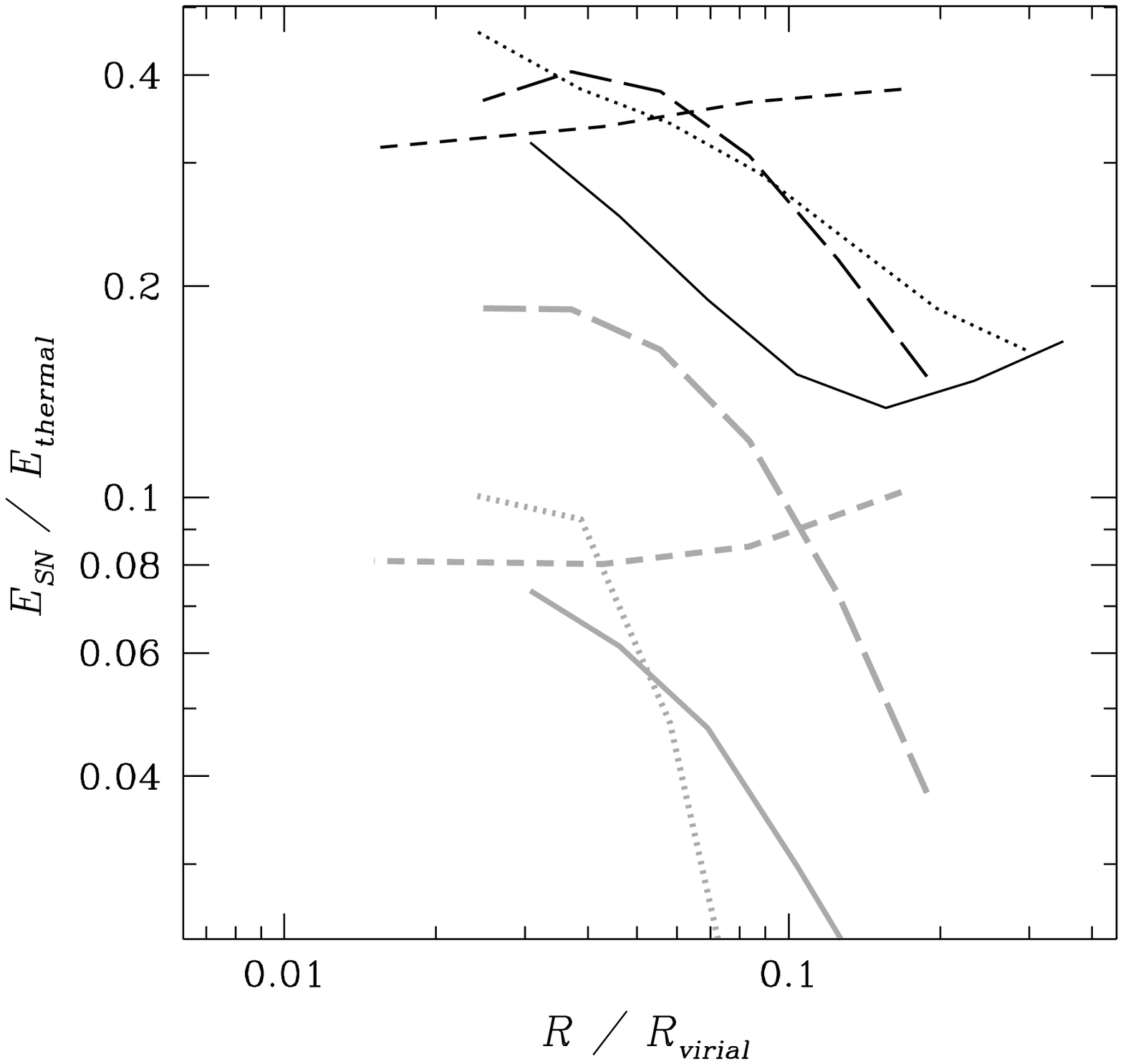}}}

\end{minipage}
\vspace*{-20pt} 
\caption{
Thermal energy associated with SNe {\it (left panel)} and relative to the
thermal energy of the gas {\it (right panel)}. Line coding is similar to
Fig.\ref{all-imlr}. Contributions of SN~Ia are shown in grey. }
\label{all-ensn}
\vspace*{-0.2cm}
\end{figure*}

\subsection{Supernovae, energy balance and gas fraction}

On the assumption that the metals measured in our systems are direct products
of SNe activity we can calculate the total energy
supplied by SNe and estimate their role in the energy balance of these
systems. 

In Fig.\ref{all-ensn} we present a comparison of the energy released by
SNe, $E_{sn}$, versus the thermal energy of the gas, $E_{th}$.  Units
chosen are keV, and the SNe values were obtained by finding the number of
SN~Ia and SN~II needed to reproduce the measured Fe and Si masses (which
are in turn derived from local metallicities) and assuming that each SN
releases $10^{51}$ ergs which is completely mixed with the gas.

Whilst SN~Ia and SN~II contribute comparable amounts of Fe from our
analysis, the much larger iron yield of SN~Ia (0.744\msun\ of released
Fe, compared to 0.121\msun\ in SN~II) implies that the total {\it
number} of SNe, and hence the total energy release, is dominated by SN~II.
This can be clearly seen in the figure.

A number of uncertainties affect our estimates of energy injection.
Firstly we have assumed that all the SN~II contribution is injected
directly into the IGM by supernovae. As discussed above, any stellar wind
material which manages to escape from galaxies will appear primarily
within our `SN~II' component, yet its injection is likely to be a less
energetic process. This, together with the fact that some of the
supernova energy will be radiated rather than lost as kinetic energy,
means that our calculated energy injection may be regarded as an upper
limit. We also neglect the energy expended in releasing the gas from
galaxy's potential well. On the other hand, the gas is released into the
system with an additional energy due to the velocity dispersion of
galaxies, which more than compensates for this loss. The sub-solar
abundances found in our systems argue in favour of strong dilution of the
galactic wind by primordial gas, which will effectively thermalise
the kinetic energy of galaxy ejecta.

In addition, the uncertainties in SN~II yields discussed above leads to
corresponding uncertainties in energy considerations. Moreover, as is
shown by the calculations of W95, SN~II from low metallicity progenitors
provide a similar Si/Fe ratio, but a much smaller mass of the released
metals, thus introducing a scaling factor to the SN~II associated energy.
In the present calculation we adopt the metal yields of the T95 model,
which assume solar metallicity progenitors. If a lower metallicity
population were assumed, the inferred SN~II energy would increase.

The main points to note from Fig.\ref{all-ensn} are the low contribution
of SN~Ia to the energy budget of the IGM, the significant contribution
(around 20--40 per cent) of SN~II, and the fact that the SNe-released energy is
more important in the poorer systems. Since the effect of significant
wind injection is primarily to reduce the gas density, by raising its
entropy (Cavaliere, Menci \& Tozzi 1997; Ponman, Cannon \& Navarro
1999), these results support the picture whereby the steepening of the
$L:T$ relation at low temperatures is a fossil remnant of the activity
generated by active star formation in galaxies (Ponman \etal 1996).

\begin{figure}
\vspace*{-0.5cm}
\centerline{\scalebox{0.5}{\epsfbox{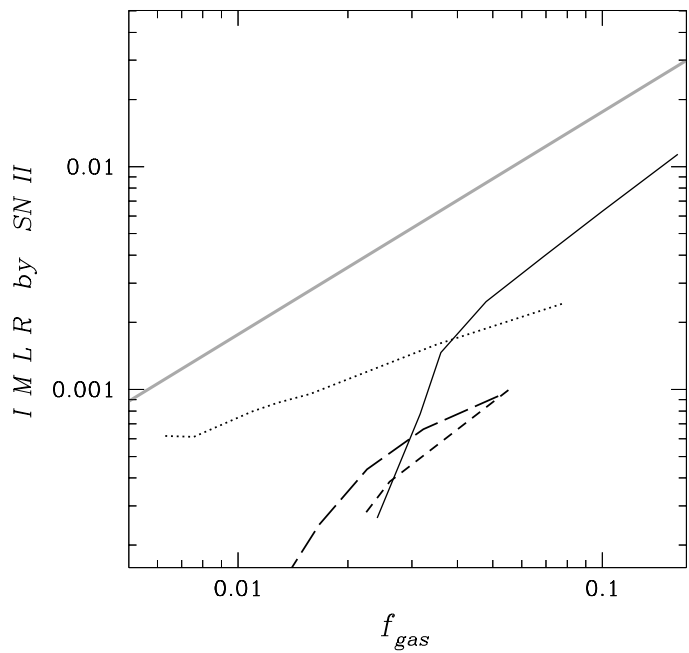}}}
\vspace*{-20pt} 
\caption{
Gas fraction vs IMLR contributed by SN~II. Line coding is similar to
Fig.\ref{all-imlr}. The expected behaviour of systems with complete mixing 
of the SN ejecta with primordial gas is shown by a thick grey line.}
\label{fgas-imlr}
\vspace*{-0.2cm}
\end{figure}

Considering the energy release by SN~II, we should take into account the
effects of possible metal loss from the system, which is suggested by the
low IMLR in the three groups compared to clusters.  If gas is lost pro-rata,
along with the metals, then the ratio of SN energy to thermal energy (which
is our main concern here) will be unaffected.  However, it is possible that
hot enriched material might escape preferentially -- for example if the hot
SN-driven wind is not completely mixed with the primordial gas and escapes
without pushing all the metal-poor gas away or some of the galaxies expelled
elements {\it before} collapsing into the primordial gas of groups or
clusters, thus bringing light but not metals. We can explore the possibility
of gas loss by comparing the observed IMLR and gas fractions ($f_{\rm gas}$)
with those of rich clusters, in which it is reasonable to suppose that gas
loss is minimal. In Fig.\ref{fgas-imlr}, we plot the relationship between
IMLR and the cumulative gas fraction $f_{\rm gas}$ (calculated 
as $M_{gas}(<r)/M_{total}(<r)$) within our observed systems.
Taking representative values within the main bodies of clusters to be an
IMLR of 0.02 (Arnaud \etal 1992) and gas fraction value of 0.17 (\eg\ Evrard
1997), we draw a line IMLR$_{\rm SN~II} \propto f_{\rm gas}$ passing through
this point. Loss of well mixed gas would move a system down this line,
whereas preferential loss of metal rich gas, or accretion of metal-poor gas,
would cause it to fall below the line.

It appears, by investigating the Fig.\ref{fgas-imlr}, that distributions of
the elements and the gas are correlated and so the changes in gas mass
fraction within one system could be explained by energy release in SN II
explosions. Although, systems demonstrate different amount of metals at a
given gas mass fraction, the difference between the groups is more then the
difference between the cluster and the groups. It might be that the age of
the group formation is different among the systems studied here, with HCG51
and NGC5044 being the youngest. We note that the behaviour of HCG62 and
NGC5044 at large radii, where the IGM gains more gas than metals, is
consistent with the idea that primordial gas may have accreted onto the
system (cf Brighenti \& Mathews, 1998).

\section{Conclusions -- supernovae and hierarchical clustering}

We have derived temperature and abundance profiles for a set of groups of
galaxies, NGC5044, HCG51 and HCG62, and compared the results with a poor
cluster AWM7. A decrease in abundances with radius is detected in all but
HCG51, which we suggest may have undergone recent disruption of its IGM,
probably due to a major merger. In AWM7 we find also a flattening of the
abundance gradients of Ne, Si, S and Fe outside a radius of 200~kpc,
correlated with a change in the abundance ratio favouring a high Si/Fe
ratio in the outer parts of the cluster. 

We have compared the derived element abundances with the predictions of
several theoretical models and Si and Fe abundance measurements as most
consistently modeled in SN~Ia -- SN~II reference frame. Employing T95 model
for the average stellar SN~II yields and TNH93 model for SN~Ia yields, we
use the spatially resolved profiles of iron and $\alpha$-elements to
decompose the iron profile into separate contributions from SN~Ia and SN~II.
Having explored their behaviour, derived the IMLR, and compared the expected
SN energy with the thermal energy of the gas, we derive the following
conclusions:

\begin{itemize} 

\item SN~Ia and SN~II have both made a significant contribution to
the enrichment of the intergalactic medium. The contribution from SN~Ia
reduces the need to invoke a flat IMF in order to explain the total mass
of heavy elements found in the IGM (Wyse 1997).

\item SN~II products are distributed widely within the IGM, which probably
indicates that they were released at early times in cluster formation via
energetic galaxy winds. Clusters have retained the SN~II products more
effectively than groups at all radii from 0.05 to at least 0.4 of
$R_{virial}$.

\item The distribution of iron abundance attributed to SN~Ia is centrally
peaked, with a core radius comparable to the optical radius of the system.
This indicates the dominance of gas release mechanisms which are strongly
density-dependent, such as gas ablation, dark halo stripping and
galaxy-galaxy interactions. Significant quantities of SN~Ia products also
appear to have been lost from the shallower potential wells of groups,
suggesting the importance of energetic injection at late times, presumably
through SN~Ia-driven winds.

\item The central rise in the $\alpha$-element abundances seen in AWM7 
is probably a product of mass loss from the cD galaxy.

\item While the SN~Ia energy is small compared to the thermal energy
of the gas, the SNe~II contribution could amount to $\sim20-40$ per cent,
which would have a substantial effect on the structure of the intracluster
medium in low mass systems.

\end{itemize}

\section*{Acknowledgments}

AF wishes to thank Maxim Markevitch, Marat Gilfanov and Eugene Churazov
for their continuous help and suggestions on the software development for
ASCA data reduction. AF acknowledges the hospitality of the University of
Birmingham and the Integral Science Data Centre (Switzerland) during the
preparation of this work, which was initiated at the International
Symposium dedicated to the third ASCA Anniversary. We would also like to thank
the referee for a careful reading of the manuscript.


\beginrefs

\BIB {Anders E. and Grevesse N. 1989, Geochimica et Cosmochimica Acta,
53, 197}
\apj{477}{128}{97}{Arimoto N., Matsushita K., Ishimaru Y., Ohashi T.,
Renzini A} 
\aa{254}{49}{92}{Arnaud M. \etal }
\apj{283}{33}{84}{Beers T.C., Geller M.J., Huchra J.P., Latham D.W., Davis R.J.}
\BIB{Burke B.E., Mountain R.W., Harrison D.C., Bautz M.W., Doty J.P., Ricker
G.R. and Daniels P.J. 1991, IEEE Trans., EED-38, 1069}
\nature{517}{311}{84}{Blumenthal G.R., Faber S.M., Primack J.R., Rees M.J.}
\apj{495}{239}{98}{Brighenti F., Mathews W. G.}
\apj{286}{408}{84}{Carlberg R.G.}
\apj{484}{L21}{97}{Cavaliere A., Menci N. and Tozzi P.}
\apj{471}{673}{96}{Churazov E., Gilfanov M., Forman W., Jones C.}
\apj{428}{544}{94}{David L., Jones C., Forman W. and Daines S.}
\apjs{110}{1}{97}{De Carvalho R.R., Ribeiro A.L.B., Capelato H.V. and
Zepf S.E.}
\apj{376}{23}{91}{Eyles \etal}
\apjl{490}{L33}{97}{Ezawa H. \etal}
\aj{100}{1}{90}{Ferguson H.C. and Sandage A.}
\BIB{Finoguenov A., Jones C., Forman W. and David L. 1999, ApJ, April issue,
(also astro-ph/9810107).}
\pasj{48}{395}{96}{Fukazawa \etal}
\BIB{Fukazawa 1997, Ph.D. Thesis, Univ. of Tokyo}
\pasj{50}{187}{98}{Fukazawa Y. \etal}
\mnras{290}{623}{97}{Gibson B.K., Loewenstein M. and Mushotzky R.F.}
\apj{255}{382}{82}{Hickson P.}
\apj{399}{353}{92}{Hickson P., Mendes de Oliveira C., Huchra J.P., Palumbo
G.G.C.} 
\astph{9710289}{97}{Hickson P.}
\apj{276}{38}{84}{Jones C. and Forman W.}
\apj{208}{177}{76}{Lampton M., Margon B. \& Bowyer S.}
\apjl{438}{L115}{95}{Liedahl D.A., Osterheld A.L. and Goldstein W.H.}
\apj{466}{695}{96}{Loewenstein M. and Mushotzky R.F.}
\apj{474}{84}{97}{Markevitch M. \& Vikhlinin A.}
\aj{97}{42}{89}{Mathews W.G.}
\BIB{Matsushita K., Makishima K., Ikebe Ya., Rokutanda E., Yamasaki
N., Ohashi T., 1998, ApJ (Letters), 499, 13}
\aa{304}{11}{95}{Matteuci F. \& Gibson B.K.}
\aas{62}{197}{85}{Mewe R., Gronenschild E.H.B.M. and Oord G.H.J.}
\BIB{Mewe R. \& Kaastra J. 1995, Internal SRON-Leiden report}
\apj{456}{80}{96}{Mulchaey J. S., Davis D.S., Mushotzky R.F., Burstein D.}
\apj{466}{686}{96}{Mushotzky R.F., Loewenstein M., Arnaud K.A., Tamura
T., Fukazawa Y., Matsushita K., Kikuchi K. and Hatsukade I.}
\aa{301}{865}{95}{Neumann D.M. \& Boehringer H.}
\nature{363}{51}{93}{Ponman T.J. and Bertram D.}
\mnras{283}{690}{96}{Ponman T.J. Bourner P.D.J., Ebeling H. and
Boehringer H.}
\nature{}{in press}{99}{Ponman T.J., Cannon D.B. and Navarro J.F.}
\BIB {Press W.H., Teukolsky S.A., Vetterling W.T., Flannery B.P. 1992,
Numerical recipes in FORTRAN}
\apjs{35}{419}{77}{Raymond J. and Smith B.}
\apj{419}{52}{93}{Renzini \etal}
\aj{104}{1039}{92}{Schweizer F. and Seitzer P.}
\apj{424}{714}{94}{Snowden S. \etal}
\BIB{Takahashi \etal 1995 ASCA Newsletter, no.3 (NASA/GSFC)}
\pasj{48}{671}{96}{Tamura T., Day C.S., Fukazawa Y., Hatsukade I., Ikebe Y.
\etal }
\pasj{46}{L37}{84}{Tanaka Y., Inoue H. and Holt S.S.}
\BIB{Thielemann, F.-K., Nomoto, K. \& Hashimoto, M. 1993 Origin and
Evolution of the Elements. (ed. N.Prantzos, E.Vangioni-Flam \& M. Cass\'e).
Cambridge Univ. Press, 297}
\apj{460}{408}{96}{Thielemann, F.-K., Nomoto, K. \& Hashimoto}
\asr{2}{241}{83}{Truemper J.}
\mnras{277}{945}{95}{Tsujimoto T., Nomoto K., Yoshii Y., Hashimoto M.,
Yanagida S. and Thielemann F.-K.}
\apj{398}{69}{92}{Worthey G., Faber S.M. and Gonzalez J.J.}
\apjs{101}{181}{95}{Woosley S.E. and Weaver T.A.}
\apj{490}{L69}{97}{Wyse R.F.G.}
\apj{462}{266}{96}{Yoshii Y., Tsujimoto T., Nomoto K.}

\endrefs

\bsp

\label{lastpage}

\end{document}